\documentclass{aa}
\usepackage{lineno}
\nolinenumbers
\usepackage{graphicx}
\usepackage{amsmath}
\usepackage{txfonts}
\usepackage{booktabs}
\usepackage{siunitx}
\usepackage{subcaption}
\usepackage{xcolor}
\usepackage{hyperref}

\begin{document}

\title{Feature redundancy, redshift, and class-prior effects in
SDSS--WISE photometric classification}

\titlerunning{Photometric classification of SDSS--WISE sources}

\author{Mohammad Sadeghi\inst{1,2}
        \and Kayoomars Karami\inst{2}}
\authorrunning{Sadeghi \& Karami}

\institute{Centre for Mathematical Plasma Astrophysics, KU Leuven, Belgium\\
           \and
           Department of Physics, University of Kurdistan, Pasdaran Street, P.O. Box 66177-15175, Sanandaj, Iran}

\date{Received ...; accepted ...}

\abstract
{Photometric classification of galaxies, quasars, and stars is important for future imaging surveys, but reported performance depends on the engineered features, spectroscopic redshift, class priors, and the sample used for evaluation.}
{We separate these four factors on 3.65 million SDSS DR17--WISE
sources, train five classifiers on twelve feature scenarios (ten reported),
and test them on 230,235 DESI DR1 sources and on 11,514 common objects.}
{We combine a four-method feature ranking, raw-only baselines, grouped
importance, and a paired bootstrap, with explicit priors, magnitude
shifts, and WISE/redshift ablations.}
{With spectroscopic redshift the tuned Random Forest reaches a macro-$F_1$
of $0.9834$; without it, 10 raw features give $0.9507$ and 24 features
$0.9551$. Redshift accounts for $80.1\%$ of the permutation importance but
mixes physics with selection: GALAXY recall at $1.5\le z<2$ is $0.016$ with
redshift and $0.68$ without. An internal photo-$z$ changes macro-$F_1$ by
only $\pm0.0001$ (necessarily redundant); an external one adds $+0.006$ to
$+0.009$ (a naive use reports $+0.015$, $62\%$ selection effect). The
no-redshift model is $8.2$ times more sensitive to magnitude shifts; QSO
purity at $0.90$ completeness falls from $0.995$ to $0.962$ at a $3\%$
quasar fraction. On DESI the QSO $F_1$ is $0.105$. A paired bootstrap on
the 11,514 common objects resolves the photometric-system effect at
$-0.05$ (95\% CI $-0.09$ to $-0.01$) for the 10-feature model, but not the
label-definition effect ($-0.02$). A major transfer failure, and in this
sample the largest single one, is the WISE depth of faint sources; this
test is confounded with optical depth and redshift, so it does not
establish that the classifier itself is not deficient. The near-zero
Pearson correlation for $W1-W2$ ($r=0.004$) is misleading: the Spearman
correlation peaks at $0.86$ in $13.0\le W1<13.5$ and falls to $0.50$ in
the faintest common bin ($14.3\le W1<17.8$), while the robust scatter of
the colour difference rises from $0.038$ to $0.087$ mag over the same
range; restricting to SDSS-range WISE photometry raises the QSO $F_1$
from $0.104$ to $0.288$.}
{Engineered features are negligible with redshift; without it, gains
come entirely from colours. Redshift is a survey-dependent signal. Internal
photo-$z$ is redundant; external gains are modest and conditional. We
recommend reporting purity/completeness under explicit priors, per-class
recall versus redshift, WISE/redshift ablations, and independent-survey
tests. Results are bounded by SDSS--WISE selection; deeper surveys
require recalibration or retraining.}

\keywords{methods: data analysis -- methods: statistical -- techniques:
photometric -- surveys -- galaxies: general -- quasars: general --
stars: general}

\maketitle
\nolinenumbers

\section{Introduction}
\label{sec:intro}

The classification of astronomical sources into galaxies, stars, and
quasars has deep historical roots, from Herschel's star gauges
\citep{herschel1789} and Hubble's identification of Andromeda as an
external galaxy \citep{hubble1929} to the recognition of quasars as
high-redshift radio sources \citep{schmidt1963} and the unification of
Seyfert galaxies with active galactic nuclei \citep{antonucci2012}. The
point-like appearance of quasars makes their morphological separation
from stars difficult; their blue optical continuum \citep{oke1963}
provides the UV excess on which the colour-based separation relies
(Sect.~\ref{sec:ablation}).

Wide-field surveys such as SDSS \citep{aguado2019} and WISE
\citep{wright2010} have catalogued hundreds of millions of sources, and
the Rubin Observatory \citep{ivezic2019} and the Roman Space Telescope
\citep{spergel2015} will increase this volume to billions. Spectroscopy
will cover only a small fraction of these sources, so photometric
classification has become a methodological necessity. At this scale, manual
labelling and even citizen-science projects such as Galaxy~Zoo
\citep{lintott2011} cannot keep pace with the data rate.

The physical basis is that multiband photometry preserves the overall shape
of the spectral energy distribution: stars show an approximately blackbody
spectrum, galaxies the composite spectrum of their stellar populations, and
quasars the relatively flat continuum of the active nucleus.

A raw magnitude mixes distance and luminosity and is therefore a weak
class indicator; at fixed redshift, a colour depends mainly on the shape
of the spectrum and is largely independent of distance, while the
residual redshift dependence (the k-correction) is itself a useful
feature (Sect.~\ref{sec:redshift}). This motivates the optical colours
($u-g$, $g-r$, $r-i$, $i-z$, $u-z$) and the WISE colours ($W1-W2$,
$W2-W3$, $W3-W4$, $i-W1$, $z-W1$, $r-W2$; throughout this paper $z$
denotes the redshift except in colour indices such as $i-z$ or $z-W1$,
where it denotes the $z$ band), following the standard SDSS indices
\citep{fukugita1996, lupton2002} and the WISE-based selection of
\citet{peters2015}. Among the colours, $W1-W2$ is a standard AGN
criterion \citep{stern2012}, because the accretion disk and dusty torus
produce a mid-infrared excess in the $3$--$5\,\mu$m range that inactive
galaxies and stars lack. As we show in Sect.~\ref{sec:ablation},
$W1-W2$ is ranked third in the combined feature ranking and second in SHAP,
but its direct ablation reveals that it represents a correlated
mid-infrared cluster rather than an independent source of information.

Feature selection in photometric classification has often relied on a
single ranking method, usually the random-forest Gini importance
\citep[e.g.,][]{bai2019, clarke2020}. Impurity-based criteria favour
features with many distinct values and can understate correlated
features \citep{strobl2007}, whereas linear tests such as the ANOVA
$F$-statistic cannot detect the nonlinear relations that govern several
physical diagnostics. We therefore combine the random-forest importance, the
mutual information, the ANOVA $F$-statistic, and SHAP values into a four-method
unified ranking, using a normalised mean, and we verify the stability of the
combined ranking with respect to the random seed and to the resampling of the data.

A large body of work applies machine learning to this problem, from
artificial neural networks \citep{bazarghan2008} and convolutional
networks \citep{shi2019, zheng2020} to semi-supervised approaches
\citep{fraixburnet2021, asadi2025}. For image-based classification, deep
convolutional networks \citep{kim2016}, single-shot object detectors
\citep{gonzalez2018}, and mask-based region networks \citep{burke2019} have
been used to classify source images. \citet{sadeghi2021} combined a support
vector machine with a genetic algorithm for feature selection and
dimensionality reduction, reaching high accuracy in the morphological
classification of galaxies.

Most closely related to this work, \citet{bai2019} trained a random
forest on nine optical-to-mid-infrared colours (SDSS, WISE, and 2MASS)
for about three million spectroscopically classified sources, reaching
accuracies above $99\%$ for stars and galaxies and above $94\%$ for
quasars. \citet{clarke2020} used 3.1 million SDSS--WISE sources to
train a random-forest classifier on the optical and mid-infrared bands
plus a morphological extension measure, and found that the WISE bands
are particularly important for quasar classification.
\citet{lourens2024} combined sharpened dimensionality reduction with a
$k$-NN classifier, reaching $F_1$ scores of $0.986$ (stars), $0.987$
(galaxies), and $0.900$ (quasars); the quasar class remains the
hardest in every study. \citet{asadi2025} combined K-means clustering
with a random-forest classifier in a semi-supervised framework, using
only 50 labelled instances to reach $F_1$ scores of $98.8\%$ (stars),
$98.9\%$ (galaxies), and $92.0\%$ (quasars) on the CPz catalog, close
to the fully supervised benchmark. \citet{chatterjee2025} compared ten
machine-learning algorithms on SDSS DR17 and found that redshift
dominates the classification when available ($64.7\%$ of the feature
importance), but that XGBoost, Random Forest, and recurrent neural
networks remain robust when redshift is excluded.

A common feature of these studies is that they either use a single
algorithm or combine several by simple voting, often without a
systematic comparison against raw-only baselines and multi-method
rankings; this work addresses that gap for tree-based classifiers on
SDSS--WISE. We therefore ask: (i) do composite features add class
information beyond raw magnitudes and the Petrosian radius? (ii) how
stable is the ranking to the seed, resampling, and method? (iii) how
much of the performance depends on redshift, how much can re-selection
recover, and is the dependence physical or a selection effect? (iv) is
a photo-$z$ from the classifier's own features independent or
redundant? (v) how do scores change with magnitude, prior, and missing
WISE data? (vi) how does an SDSS model perform on DESI? Our main
results are: engineered features are negligible with redshift and add
only $\approx0.4$ percentage points without it, almost entirely from
colours; redshift is the most important feature but its importance
mixes physics (the stellar peak at $z\approx0$) with the SDSS
selection (the near-absence of galaxies above $z\approx1$); and an
SDSS classifier does not transfer to DESI without prior correction and
photometric adaptation. A major transfer failure, and in this sample
the largest single one, is the WISE depth of faint sources; this test is
confounded with optical depth and redshift, so it does not establish that
the classifier itself is not deficient. We use the raw
SDSS--WISE catalogue without per-class outlier removal or
label-dependent preprocessing, so all metrics are computed on the test
set without class-dependent filtering. All conclusions are established
for the SDSS--WISE selection function; extending them to deeper
surveys requires retraining or recalibration on the target photometric
system. Sections~\ref{sec:data}--\ref{sec:discussion} present the data, features, classification
performance, diagnostic tests, DESI validation, and discussion;
supplementary material is in Appendix~\ref{app:supplementary}.

\section{Data and preprocessing}
\label{sec:data}

\subsection{Query and cross-matching}
\label{sec:query}

We obtained the SDSS DR17 data \citep{abdurrouf2022} by querying the
CasJobs service of the SkyServer API\footnote{\url{https://skyserver.sdss.org/CasJobs/}},
joining the \texttt{SpecObj} and \texttt{PhotoObj} tables for all
sources with a spectroscopic class (GALAXY, QSO, or STAR), a redshift
above $-0.1$ (to retain nearby stars with negative radial velocities),
and a positive $u$-band magnitude. The spectroscopic class and redshift
are measured by the SDSS spectroscopic pipeline \citep{bolton2012}, the
photometric quantities by the imaging pipeline \citep{stoughton2002}.
We retrieved the object identifier, sky coordinates, class and redshift,
the five model magnitudes ($u,g,r,i,z$), and the $r$-band Petrosian
radius (\texttt{petroRad\_r}). Each source was cross-matched by position
with the WISE All-Sky catalog through the SDSS-provided
\texttt{wise\_xmatch} table to obtain profile-fit (point-source)
magnitudes in the four WISE bands ($W1$--$W4$) \citep{wright2010}. The
identifier and sky coordinates are not used as features; we also
computed a binary source-extension flag
($|m_{\rm petro,r}-m_{\rm psf,r}|>0.2$ mag) but did not use it because
it is largely redundant with the continuous Petrosian radius. The
redshift is the spectroscopic redshift measured from the same spectrum
that defines the class; it is not a photometric redshift, and because it
is measured from the same spectrum that defines the label it can act as a
class-prior shortcut (Sect.~\ref{sec:redshift}). The query returned
\num{4639848} unique sources: \num{2811139} ($60.6\%$) galaxies,
\num{866443} ($18.7\%$) quasars, and \num{962266} ($20.7\%$) stars.

\subsection{Missing data and selection effects}
\label{sec:missing}

No optical magnitude, redshift, or Petrosian radius is missing. WISE
photometry is unavailable for \num{987538} sources ($21.3\%$), either
because the source has no WISE counterpart or because at least one of
the four WISE bands has a non-detection sentinel value ($-9999$). All
such sources were removed, leaving \num{3652310} sources with valid
WISE photometry in all four bands. The loss depends strongly on the
class: $10.3\%$ of galaxies, $33.7\%$ of quasars, and $42.1\%$ of
stars (the three figures sum to 986,650, within $0.1\%$ of the total).
We tentatively attribute this to the limited depth of WISE, whose
$5\sigma$ point-source sensitivities are about $0.08$, $0.11$, $1$,
and $6$\,mJy in $W1$--$W4$ \citep{wright2010}: faint quasars and
distant stars are expected to fall below the limit more often than
bright, nearby galaxies. We have not tested this explanation with the
WISE depth maps. Requiring valid WISE photometry changes the class
fractions of GALAXY/QSO/STAR from $60.6/18.7/20.7\%$ to
$69.0/15.7/15.3\%$. This selection limits the sample to which our
conclusions apply, and it is relevant for surveys whose WISE coverage
differs.

\subsection{Exploratory analysis}
\label{sec:eda}

In the optical bands, galaxies and quasars overlap strongly, and stars
are offset toward brighter magnitudes. In $W1$ and $W2$ the three
classes also overlap, and in $W3$ and $W4$ the densities are nearly
flat, consistent with fluxes close to the WISE detection limit. Only
the redshift and the Petrosian radius separate the classes visually.
The $r$-band distribution of galaxies has maxima near $r\approx17.6$
and $r\approx20.5$; the brighter one lies at the $r<17.77$ limit of
the SDSS main galaxy sample.

The redshift distributions differ by orders of magnitude, so we
describe them separately. Stars form a narrow peak at zero
($|z|\lesssim10^{-3}$, radial velocities of at most a few hundred
km\,s$^{-1}$), as expected for sources without cosmological redshift.
Galaxies are bimodal, with maxima near $z\approx0.1$ and
$z\approx0.5$; we suspect that this reflects two target-selection
subsamples of the SDSS spectroscopic catalog but have not checked
this against the target-selection bitmasks. Quasars have a broad
unimodal distribution that peaks near $z\approx1.2$ and extends beyond
$z\approx6$. Stars are therefore almost disjoint from the other
classes in redshift, whereas galaxies and quasars overlap at
$z\lesssim1$: the galaxy distribution ends near $z\approx1$ and only
a few hundred galaxies lie at $1.5\le z<2$. This sharp end of the
galaxy sample is important for the redshift analysis
(Sect.~\ref{sec:redshift}), because a model that has access to $z$ can
learn that $z\gtrsim1$ means ``quasar''.

\subsection{Split and preprocessing}
\label{sec:split}

The \num{3652310} WISE-detected sources are split $70/15/15$ into
training (\num{2556617}), validation (\num{547846}), and test
(\num{547847}) sets, stratified by class, before any step that depends
on the class distribution. The pipeline contains no per-class outlier
removal and no label-dependent preprocessing: no feature construction,
transformation, or filtering step uses the true class of the source.
Benchmark studies that filter the test set with the true class can
report higher scores; we have not quantified this effect for the
present sample.

\section{Feature engineering and ranking}
\label{sec:features}

\subsection{Composite features}
\label{sec:composite}

From the 11 raw features ($u,g,r,i$, the $z$-band magnitude $m_z$,
redshift, $R_{\rm petro}$, and $W1$--$W4$) we construct 14 additional
features: eleven colours ($u-g$, $g-r$, $r-i$, $i-z$, $u-z$,
$W1-W2$, $W2-W3$, $W3-W4$, $i-W1$, $z-W1$, $r-W2$) and three
morphology features. The morphology features are the mean surface
brightness within the Petrosian radius,
$\mu_r=m_r+2.5\log_{10}(\pi R_{\rm petro}^2)$, with $R_{\rm petro}$ in
arcsec; the compactness $C=1/(R_{\rm petro}+\epsilon)$; and
$\log_{10}(R_{\rm petro}+\epsilon)$, with $\epsilon=10^{-3}$ (in
arcsec) to avoid a singularity at $R_{\rm petro}=0$. This gives a
25-feature set.

The composite features are not independent of one another. First, $C$
is a strictly decreasing function of $R_{\rm petro}$ and
$\log_{10}(R_{\rm petro}+\epsilon)$ a strictly increasing one, so a
tree-based model can represent them identically and their ranks in a
feature ranking are arbitrary. Second, the colours are linear
combinations of magnitudes; they add no information in the
information-theoretic sense (the mutual information with the label is
invariant under an invertible linear transformation). A tree splits on
one feature at a time, however, so a colour is not equivalent to its
two magnitudes for an axis-aligned model, and whether it helps is an
empirical question that we answer in Sect.~\ref{sec:ablation}. Third,
$u-z$ is exactly the sum of the four successive optical colours, and
$\mu_r$ is a function of $m_r$ and $R_{\rm petro}$.

The optical colours trace the shape of the spectral energy
distribution. The UV excess of quasars makes them bluer in $u-g$
(density peak near $0.4$, compared with $1.7$ for galaxies) and in
$u-z$ (peak near $0.6$, compared with $4$ for galaxies), and the
$4000$\,\AA\ break moves through the bands as the redshift of a galaxy
increases (Sect.~\ref{sec:redshift}). The mid-IR colour $W1-W2$ is
larger for quasars (peak near $1$) than for stars and galaxies (near
$0$ and $0.2$) because of the emission of hot dust \citep{stern2012}.
The surface brightness $\mu_r$ separates the compact stars (peak near
$19$ mag arcsec$^{-2}$) and quasars ($22$ mag arcsec$^{-2}$) from the
extended galaxies ($24$ mag arcsec$^{-2}$), while the densities of
$W2-W3$ and $W3-W4$ are nearly flat.

Because the composite features are constructed from the raw ones,
many of them are correlated by construction. Colours correlate with
their own bands (e.g., $g-r$ with $g$), $W1-W2$ and $r-W2$ share the
$W2$ term (Pearson $r\approx0.9$ on the training set), and the three
morphology features are functions of $R_{\rm petro}$ alone. The galaxy
density of $\log_{10}R_{\rm petro}$ shows narrow peaks at $\approx0.47$,
$0.87$, $1.05$, and $1.25$ ($R_{\rm petro}\approx3.0$, $7.4$, $11.4$,
and $18.2$\,arcsec), which suggests that the measured radius is
quantised. We tentatively attribute this to the discrete annuli of the
SDSS radial profiles but have not verified it.

\subsection{Multi-method importance ranking}
\label{sec:ranking}

We rank the 25 features on the training set with four methods: (i) the
mean decrease in Gini impurity of a random forest of 300 trees with
class-balanced sample weights; (ii) the mutual information with the
class label from a $k$-nearest-neighbour estimator; (iii) the ANOVA
$F$-statistic; and (iv) the mean absolute SHAP value
\citep{lundberg2017} from a LightGBM model. We use LightGBM for the
SHAP calculation because tree SHAP is exact for gradient-boosted trees;
the resulting ranking is then applied to the Random Forest model used
for classification. Each score is min--max normalised to $[0,1]$ and
the combined score is the unweighted mean. Table~\ref{tab:ranking}
gives the Top-12.

Of these 12 features, only redshift and $R_{\rm petro}$ are raw; the
other ten are engineered. This does not by itself show that the
engineered features carry information beyond the raw ones: impurity
importance, mutual information, and the $F$-test all split credit among
correlated features, and the order inside a correlated group is
arbitrary. The direct test is in Sect.~\ref{sec:ablation}. No raw
magnitude ($u,g,r,i,m_z$) is among the 12 highest-ranked features, in
line with the argument that apparent magnitudes mix luminosity and
distance.

\begin{table}[htbp]
\caption{Top 12 features by combined normalised importance score.}
\label{tab:ranking}
\centering
\begin{tabular}{rlrl}
\hline\hline
Rank & Feature & Score & Type \\
\hline
1  & redshift                    & 1.000 & raw \\
2  & $\log_{10}R_{\rm petro}$    & 0.628 & engineered \\
3  & $W1-W2$                     & 0.620 & engineered \\
4  & compactness                 & 0.456 & engineered \\
5  & $g-r$                       & 0.368 & engineered \\
6  & $r-i$                       & 0.343 & engineered \\
7  & $R_{\rm petro}$             & 0.312 & raw \\
8  & $\mu_r$                     & 0.298 & engineered \\
9  & $u-z$                       & 0.275 & engineered \\
10 & $u-g$                       & 0.241 & engineered \\
11 & $r-W2$                      & 0.229 & engineered \\
12 & $i-z$                       & 0.205 & engineered \\
\hline
\end{tabular}
\end{table}

\subsection{Stability of the ranking}
\label{sec:stability}

We repeat the ranking with ten random seeds and with 30 stratified
bootstrap resamples ($80\%$ of each class, with replacement). The
Kendall $\tau$ between the reference ranking (seed 42) and the other
nine is $0.976$ on average (range $0.967$--$0.993$). Redshift,
$\log_{10}R_{\rm petro}$, and compactness keep ranks 1, 2, and 4 in
all seeds (with $W1-W2$ taking rank 3), $W4$ is always last, and the
rank changes occur between features of the same family (morphology or
WISE). In the bootstrap, redshift is the only feature with zero rank
variance. The SHAP rank of $\log_{10}R_{\rm petro}$ is $12.2\pm8.7$,
whereas its rank from the three non-SHAP methods is $3.0\pm0.4$, as
expected for a group of near-duplicate features whose credit depends on
the resample. Table~\ref{tab:bootstrap} in
Appendix~\ref{app:supplementary} lists the stability of the main
features.

\section{Classification performance}
\label{sec:class}

\subsection{Experimental protocol}
\label{sec:protocol}

We train five classifiers (LinearSVC with Platt-sigmoid-calibrated
probabilities, Random Forest, XGBoost, LightGBM, and CatBoost) on the
full 25-feature set and on twelve feature scenarios. The scenarios are
the Top-12 and Top-6 sets from three ranking variants (the three-method
unified ranking, the four-method unified ranking, and SHAP), each with
redshift and with redshift removed from the selected set (ablation). Of
these twelve scenarios, ten are reported in
Table~\ref{tab:all_scenarios}; the two Top-12 variants whose
with-redshift and no-redshift entries both appear only in
Table~\ref{tab:three_protocol} enter only the no-redshift comparison of
that table. Six additional re-selection variants, in which the feature
selection is performed again on a $z$-free pool, are evaluated in
Sect.~\ref{sec:redshift}. The best algorithm per scenario is chosen on
the validation macro-$F_1$, and the test set is used once per scenario,
for the single selected model. Here ``12-feature ablation'' denotes a
set selected with redshift and then used with redshift removed
(11 features).

Throughout the paper we report three values of the Top-12 SHAP
macro-$F_1$: $0.9834$ for the tuned Random Forest on the test set
(Tables~\ref{tab:per_class_best}, \ref{tab:a1}, \ref{tab:all_scenarios}),
$0.9832$ for the default Random Forest on the same test set
(Sect.~\ref{sec:tuning}), and $0.9830$ for the permutation-model
baseline in Sect.~\ref{sec:grouped}. The differences reflect the tuning
and the permutation baseline, not the data split. The bootstrap
intervals in Table~\ref{tab:all_scenarios} capture the sampling
variability of the test set with the model and split held fixed; they
do not include the variability of the training seed or of the
train/validation/test split. Differences below about $0.003$ between
configurations should therefore be read as indicative rather than as
established.

Table~\ref{tab:all_scenarios} in Appendix~\ref{app:supplementary} gives
the full test-set macro-$F_1$ with bootstrap $95\%$ intervals. The
Random Forest is selected in every scenario. \emph{With redshift}, the
12-feature scenarios reach $0.9827$--$0.9834$
(Tables~\ref{tab:all_scenarios} and~\ref{tab:a1}) and the 25-feature
scenario reaches $0.9830$, so the 13 lowest-ranked features make a
practically negligible contribution. The three Top-12 selections differ
by at most $0.0007$, which is practically negligible; we do not claim
statistical equivalence, because with $N\approx5.5\times10^5$ even a
difference of $0.0004$ is formally significant in a McNemar test.
\emph{Without redshift}, the 10 raw features give $0.9507$ and the full
24-feature set reaches $0.9551$, a gain of $0.0044$; the 12-feature
ablation scenarios reach $0.9526$--$0.9531$ and the 6-feature ablation
scenarios $0.9001$--$0.9147$.

\subsection{Hyperparameter tuning}
\label{sec:tuning}

We tune three tree-based algorithms (Random Forest, XGBoost, LightGBM)
in the twelve scenarios. For the Random Forest we use a 15-candidate
random search scored by the out-of-bag macro-$F_1$. For XGBoost and
LightGBM we use a 40-candidate random search with three-fold
cross-validation followed by a targeted grid refinement of the two most
influential hyperparameters. All searches use a stratified subsample of
\num{300000} training rows; the winning configuration is refitted on
the full training set with three random seeds averaged at the
probability level.

Table~\ref{tab:tuning_gains} in Appendix~\ref{app:supplementary} shows
the effect of tuning. Tuning a Random Forest is on average neutral or
harmful: the mean test gain is $+0.0001$ and $+0.0003$ for the Top-12
scenarios with and without redshift, $-0.0009$ for the Top-6 scenarios
with redshift, and $-0.0058$ for the no-redshift Top-6 scenarios
(largest single regression $-0.0087$). XGBoost is the most tunable
algorithm, with mean gains of $+0.0030$ and $+0.0034$ for the Top-12
scenarios with and without redshift; LightGBM is intermediate
($+0.0012$ and $+0.0032$; largest single gain $+0.0066$ for the Top-12
SHAP scenario without redshift). Most changes are smaller than $0.003$,
which is comparable to the width of the bootstrap intervals
($0.0006$--$0.0019$); only gains above about $0.003$ are likely robust
to the training noise, which the bootstrap intervals do not capture. A
plausible reason for the Random Forest losses is that the search is run
on a $3\times10^5$ subsample, whereas the final model is trained on
$2.56\times10^6$ sources, so the optimum of the subsample is not
necessarily the optimum of the full data.

The learning curve of the default Random Forest (Sect.~\ref{sec:protocol})
rises from $0.9737$ at $10^4$ training sources to $0.9807$ at $10^5$
and flattens at about $5\times10^5$: increasing the training set from
$5\times10^5$ to the full $2.56\times10^6$ raises the macro-$F_1$ by
only $0.0005$ (from $0.9827$ to $0.9832$). The remaining error is
therefore not limited by the number of training sources but by the
information in the features and by the intrinsic overlap of the
classes. Training on $10^5$ sources loses only about $0.0025$.

\subsection{Per-class performance}
\label{sec:perclass}

\begin{figure*}[htbp]
\centering
\begin{subfigure}{0.42\textwidth}
  \includegraphics[width=\linewidth]{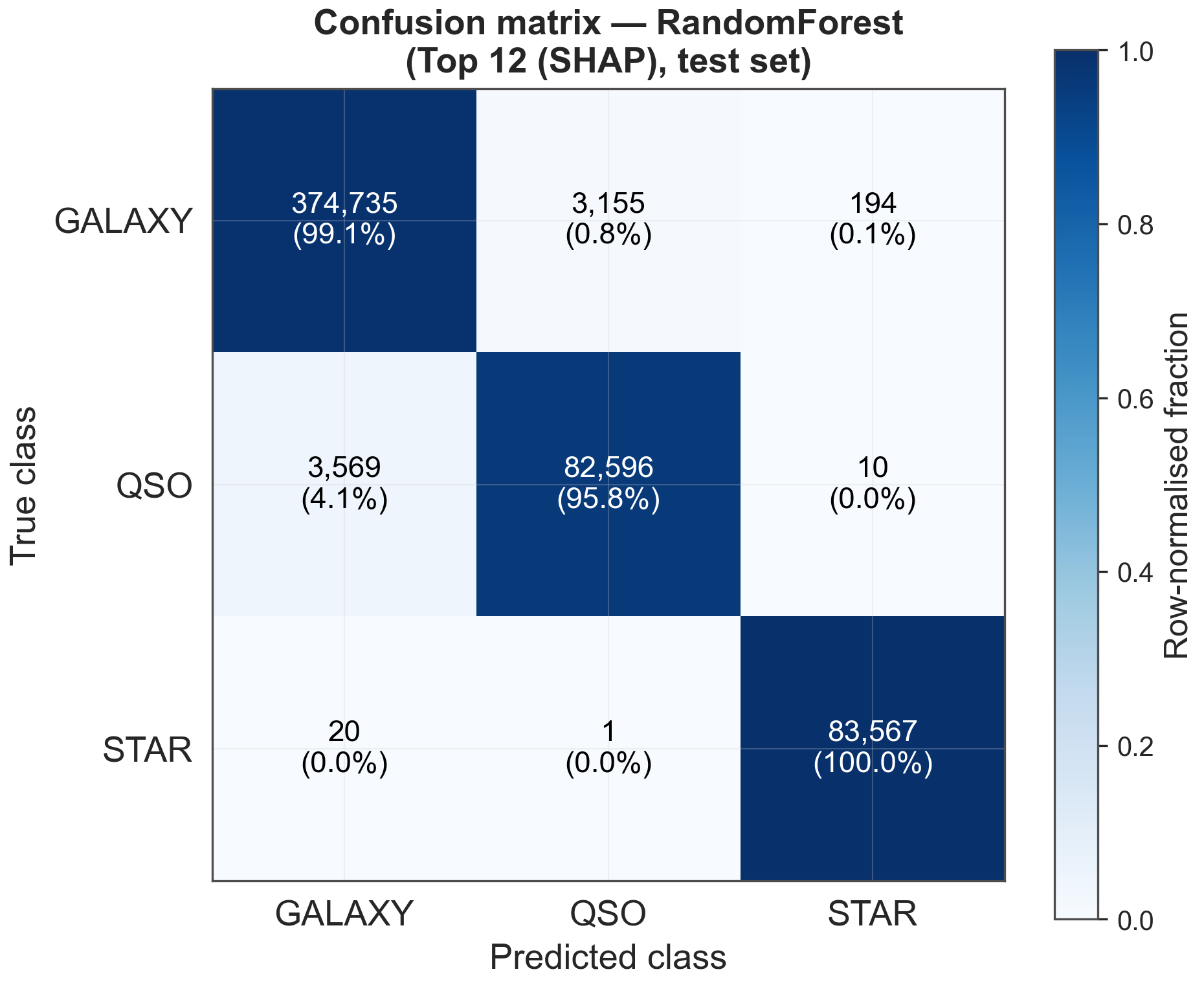}
\end{subfigure}\hfill
\begin{subfigure}{0.42\textwidth}
  \includegraphics[width=\linewidth]{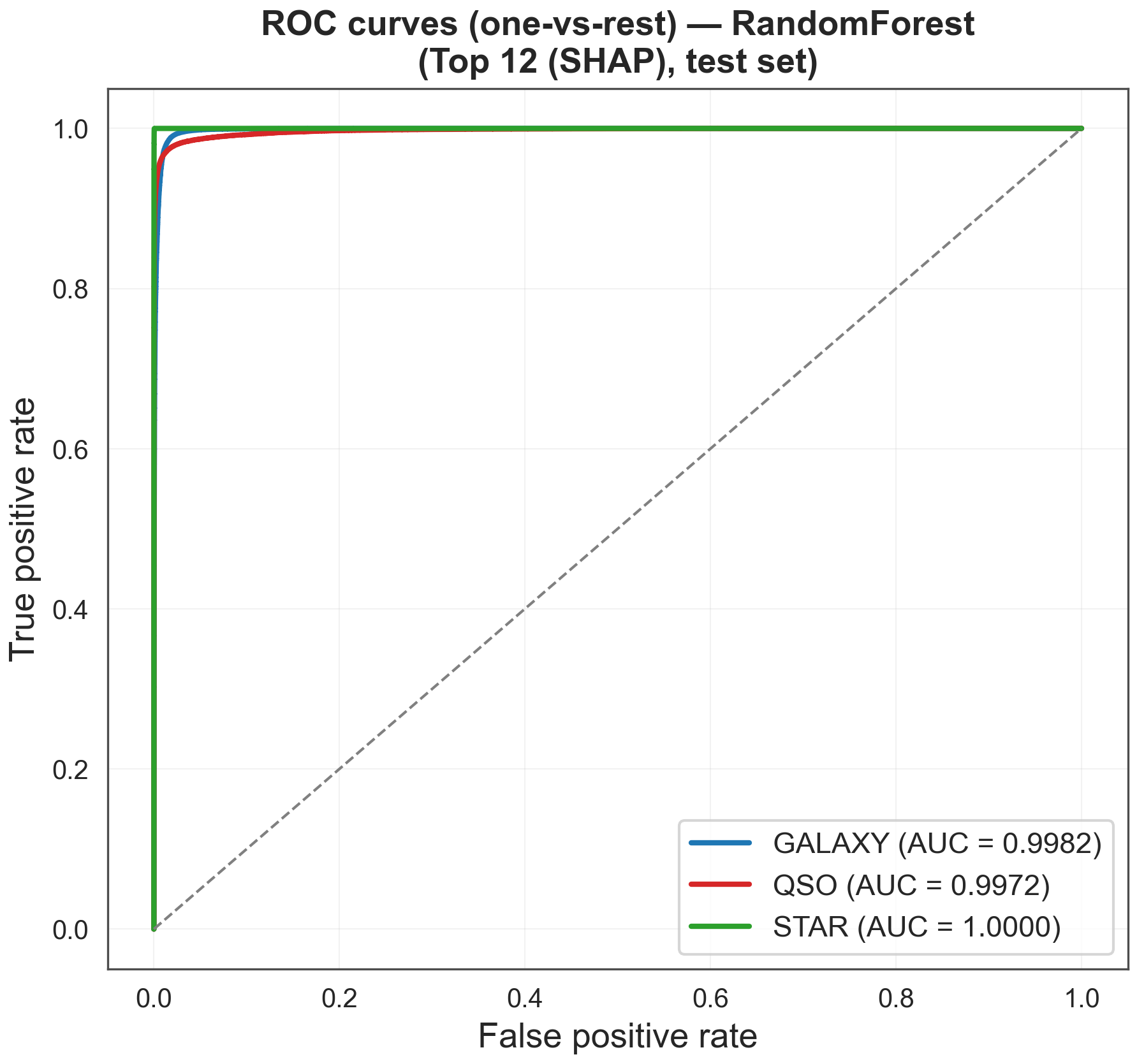}
\end{subfigure}
\caption{Test confusion matrix (left) and one-vs-rest ROC curves
(right) for the tuned Random Forest on the Top-12 (SHAP) set.}
\label{fig:cm_roc}
\end{figure*}

Table~\ref{tab:per_class_best} gives the per-class performance of the
best model (tuned Random Forest, Top-12 SHAP set, with redshift), and
Fig.~\ref{fig:cm_roc} shows the confusion matrix and the one-vs-rest
ROC curves. The star class is almost perfect ($F_1=0.9987$; only $21$
of \num{83588} stars are misclassified). Of the \num{378084} galaxies,
\num{3155} are classified as quasars and $194$ as stars; of the
\num{86175} quasars, \num{3569} are classified as galaxies and $10$ as
stars. The QSO class has the lowest $F_1$ ($0.9608$). Of the
\num{6949} errors in total, \num{6724} ($96.8\%$) are
GALAXY$\leftrightarrow$QSO confusions, and the two directions are
nearly symmetric. The one-vs-rest AUC is $0.9982$, $0.9972$, and
$1.0000$ for galaxies, quasars, and stars: the ranking of the sources
by probability is near-perfect, and the errors concentrate near the
decision boundary. The ROC curve does not depend on the class prior, in
contrast to the precision, which is the subject of
Sect.~\ref{sec:purity}.

\begin{table}[htbp]
\caption{Per-class test performance of the best model.}
\label{tab:per_class_best}
\centering
\begin{tabular}{lrrrr}
\hline\hline
Class & Precision & Recall & $F_1$ & Support \\
\hline
GALAXY & 0.9905 & 0.9911 & 0.9908 & \num{378084} \\
QSO    & 0.9632 & 0.9585 & 0.9608 & \num{86175}  \\
STAR   & 0.9976 & 0.9997 & 0.9987 & \num{83588}  \\
Macro avg & 0.9838 & 0.9831 & 0.9834 & \num{547847} \\
\hline
\end{tabular}
\end{table}

\subsection{Per-class recall as a function of redshift}
\label{sec:zbinned}

\begin{figure}[htbp]
\centering
\includegraphics[width=0.45\textwidth]{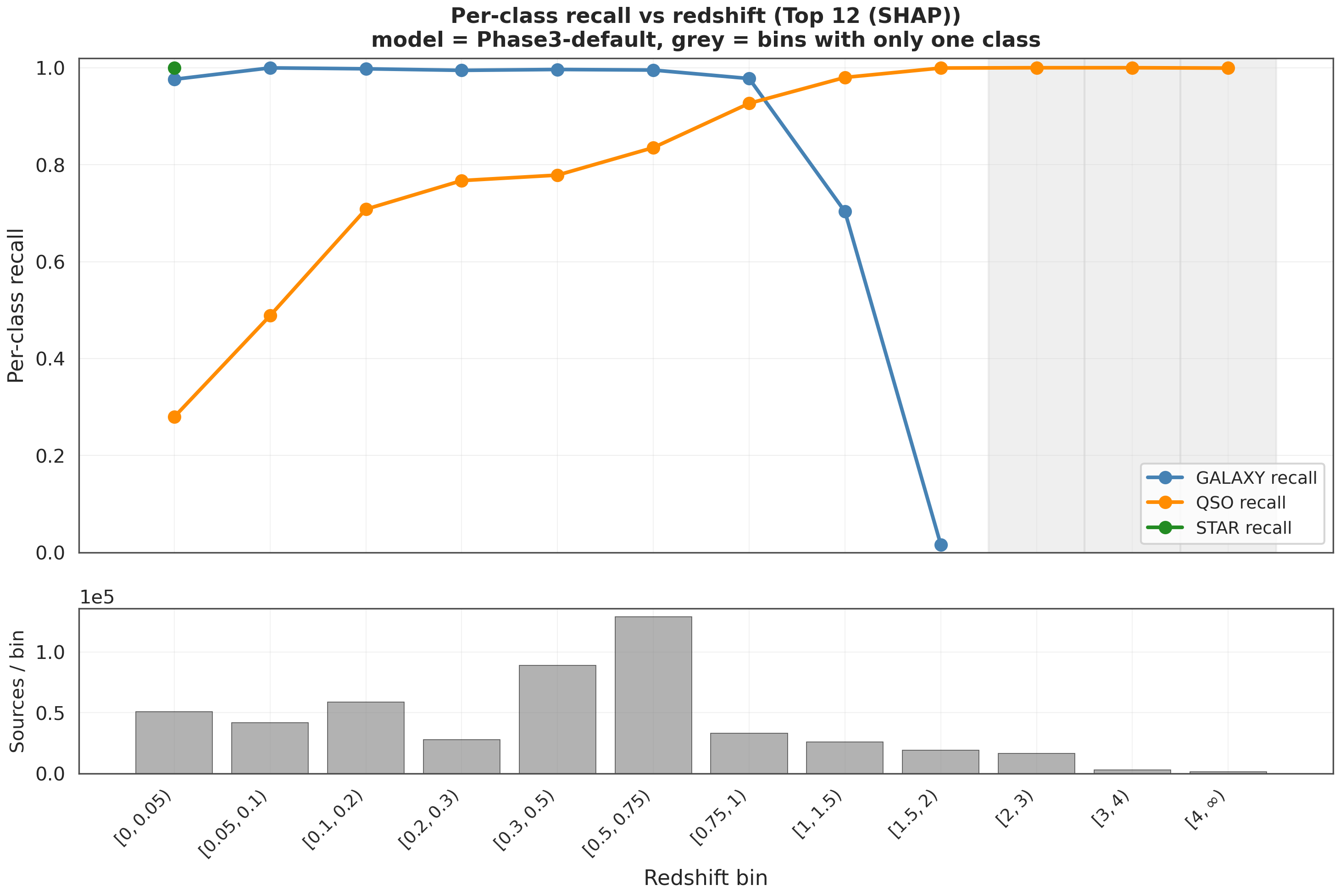}
\caption{Per-class recall as a function of redshift for the tuned
Random Forest on the Top-12 (SHAP) set. The 51,255 stars with $z<0$
are not shown because the bins start at $z=0$.}
\label{fig:redshift_recall}
\end{figure}

Fig.~\ref{fig:redshift_recall} shows the per-class recall as a function
of redshift for the best model, and Fig.~\ref{fig:recall_noz} in
Appendix~\ref{app:supplementary} the same quantity for the no-redshift
model; the numbers are listed in Table~\ref{tab:recall_binned_all} in
Appendix. The stellar class has a recall of $0.9994$ in the lowest bin
($0\le z<0.05$, \num{32333} stars); the bins start at $z=0$, so the
\num{51255} test stars with $z<0$ are not included. The GALAXY recall
is at least $0.976$ up to $z\approx1$ and falls to $0.016$ in the bin
$1.5\le z<2$, where only $306$ galaxies remain. The QSO recall is low
at low redshift ($0.28$ at $z<0.05$, $0.49$ at $0.05\le z<0.1$, $0.71$
at $0.1\le z<0.2$), rises to $0.93$ at $0.75\le z<1$, and reaches
$1.0$ above $z=1.5$. The low-redshift numbers rest on small samples of
quasars ($93$ and $264$ sources in the first two bins) and on a
physical difficulty: low-redshift quasars have a host galaxy that
contributes to the optical light, so that their colours and sizes
overlap with those of galaxies.

Without redshift (Fig.~\ref{fig:recall_noz} in
Appendix~\ref{app:supplementary}), the GALAXY recall in the lowest bin
falls from $0.976$ to $0.948$ and the STAR recall from $0.999$ to
$0.934$. The QSO recall changes only from $1.0$ to $0.978$ at
$2\le z<3$, but falls to $0.912$ at $3\le z<4$ and to about $0.39$ for
$z\ge4$ (Table~\ref{tab:recall_binned_all}). This behaviour is expected
physically: beyond $z\approx3$ the Lyman-$\alpha$ forest removes a
growing part of the flux in the $u$ and $g$ bands, so that the optical
colours of a quasar no longer resemble the blue continuum that
identifies it at lower redshift. In the opposite direction, the
no-redshift model recovers galaxies where the redshift model fails: at
$1.5\le z<2$ the GALAXY recall is $0.680$ without redshift, against
$0.016$ with it.

\subsection{Do the engineered features add information?}
\label{sec:ablation}

Table~\ref{tab:ablation} compares raw-only and engineered scenarios. With
redshift, the 11 raw features give $0.9830$, essentially the same as the
best engineered 12-feature set ($0.9834$); without redshift, the 10 raw
features give $0.9507$ and the full 24-feature set $0.9551$. To identify
which group drives this gain of $0.0044$, we add each group to the raw
features separately. The optical colours add $+0.0024$ and the WISE
colours $+0.0013$, whereas the morphology features alone do not help
($-0.0015$); combining the optical and WISE colours ($0.9553$)
reproduces the full set. The gain therefore comes entirely from the
colours; the three morphology features add no measurable gain once the
colours are present, and alone they slightly degrade the score because
they are monotonic functions of the Petrosian radius that the raw
feature already provides. By contrast, removing the WISE bands costs
$3.0$\,pp (Sect.~\ref{sec:deployability}), about seven times the gain
from all engineered features. A direct ablation of $W1-W2$ changes the
test macro-$F_1$ by $-0.0002$ in the 12-feature set (from $0.9828$ to
$0.9826$) and by $-0.0024$ in the 6-feature set (from $0.9787$ to
$0.9763$); its contribution depends on whether $r-W2$, which shares the
$W2$ term, is available. $W1-W2$ is thus a representative of a
correlated mid-infrared cluster rather than an independent source of
class information.

\begin{table}[htbp]
\caption{Raw-only versus engineered scenarios, and feature-group
ablation without redshift.}
\label{tab:ablation}
\centering
\begin{tabular}{lrr}
\hline\hline
Feature set & $n$ feat & Macro-$F_1$ \\
\hline
Raw, with $z$                       & 11 & 0.9830 \\
All features (25), with $z$         & 25 & 0.9830 \\
Top 12 (SHAP), with $z$             & 12 & 0.9834 \\
\hline
Raw, no $z$                         & 10 & 0.9507 \\
Raw + optical colours, no $z$       & 15 & 0.9531 \\
Raw + WISE colours, no $z$          & 16 & 0.9520 \\
Raw + morphology, no $z$            & 13 & 0.9492 \\
Raw + optical + WISE, no $z$        & 21 & 0.9553 \\
Raw + all engineered, no $z$        & 24 & 0.9551 \\
Top 12 (SHAP), no $z$               & 11 & 0.9526 \\
Photometric (14 feat, no $z$)       & 14 & 0.9561 \\
Optical only (no WISE)              &  8 & 0.9234 \\
\hline
\end{tabular}
\end{table}

\subsection{Grouped permutation importance}
\label{sec:grouped}

Permutation importance computed feature by feature is misleading when
features are correlated: when one member of a correlated group is
permuted, the model can substitute one of the others, and the measured
importance of that single feature understates the contribution of the
group. We therefore define clusters by physical category (redshift,
morphology size, optical colours, mid-IR colours, surface brightness)
and permute all members of a cluster jointly
(Table~\ref{tab:perm_domain}; per-feature values are in
Table~\ref{tab:perm_standard_v2}). The effect is largest for the
morphology group: the three members have per-feature importances of
$0.0001$, $0.0001$, and $0.0005$ in the with-redshift model (sum
$0.0007$), while their joint importance is $0.0021$; in the no-redshift
model the joint importance is $0.1422$. The value $0.0021$ is the
grouped importance reported in Table~\ref{tab:perm_domain}, not the sum
of the per-feature values. The increase to $0.1422$ is physical in the
sense that size is one of the few remaining ways to separate point
sources from galaxies, but it is not independent of the raw Petrosian
radius, and the ablation of Sect.~\ref{sec:ablation} shows that the
engineered morphology features add no measurable gain once the colours
are present; the $0.1422$ therefore measures how much the model relies
on size, not how much the engineered features add beyond it. Redshift
alone carries $0.5312$, $80.1\%$ of the total importance, and the
baseline macro-$F_1$ of the permutation model falls from $0.9830$ to
$0.9542$ when redshift is removed.

\begin{table}[htbp]
\caption{Grouped permutation importance of feature clusters.}
\label{tab:perm_domain}
\centering
\resizebox{\columnwidth}{!}{%
\begin{tabular}{lrccr}
\hline\hline
Cluster & $n$ feat & With $z$ & No $z$ & Ratio \\
\hline
Redshift            & 1 & $\mathbf{0.5312}$ ($80.1\%$) & --- & --- \\
Optical colours (5) & 5 & $0.0923$ & $\mathbf{0.2927}$ & $3.2\times$ \\
Mid-IR colours (2)  & 2 & $0.0190$ & $\mathbf{0.2508}$ & $13.2\times$ \\
Morphology size (3) & 3 & $0.0021$ & $\mathbf{0.1422}$ & $\mathbf{67.7\times}$ \\
Surface brightness  & 1 & $0.0182$ & $\mathbf{0.1326}$ & $7.3\times$ \\
\hline
\end{tabular}
}
\end{table}

\section{Diagnostic tests}
\label{sec:diagnostics}

\subsection{Dependence on the spectroscopic redshift}
\label{sec:redshift}

The classes are labelled from the spectra, from which the redshift is
also measured, so the redshift is not an independent photometric
quantity; for stars it is almost a label ($|z|\le 10^{-3}$). The scores
with redshift are therefore upper bounds for a photometric survey. We
compare three protocols (Table~\ref{tab:three_protocol}): (a) the
reference with redshift; (b) an ablation, in which redshift is removed
from a set selected with it; and (c) a re-selection, in which the
features are selected again from the no-redshift pool. At 12 features,
removing the redshift lowers the macro-$F_1$ by $3.0$--$3.1$ percentage
points (pp), and re-selection recovers $+0.0001$ to $+0.0024$ of it;
the 12-feature sets already contain colours that act as proxies for the
galaxy redshift (Spearman correlations $+0.89$ for $r-W2$, $+0.88$ for
$i-W1$, $+0.86$ for $r-i$, and $+0.85$ for $z-W1$; but only
$0.14$--$0.16$ for stars and between $-0.04$ and $-0.50$ for quasars;
Table~\ref{tab:spearman}). At 6 features, the ablation costs
$6.4$--$7.9$\,pp; re-selection recovers $1.8$--$5.0$\,pp, but the
macro-$F_1$ is still $2.8$--$5.3$\,pp below the reference, so a
6-feature classifier without redshift is usable only after a new
selection.

\begin{table}[htbp]
\caption{Three-protocol comparison of the no-redshift deployment scores.}
\label{tab:three_protocol}
\centering
\resizebox{\columnwidth}{!}{%
\begin{tabular}{lrrrrrrr}
\hline\hline
Scenario & $n_{\rm with}$ & $n_{\rm no}$ & With $z$ & Ablation & Re-selection & $\Delta$ & Retention \\
\hline
Top 12 (4-method unified) & 12 & 11 & 0.9828 & 0.9530 & \textbf{0.9541} & $+0.0011$ & 97.1\% \\
Top 12 (SHAP)             & 12 & 11 & 0.9834 & 0.9526 & \textbf{0.9550} & $+0.0024$ & 97.1\% \\
Top 12 (3-method)         & 12 & 10 & 0.9827 & 0.9531 & 0.9532 & $+0.0001$ & 97.0\% \\
Top 6 (4-method unified)  &  6 &  5 & 0.9783 & 0.9147 & \textbf{0.9382} & $+0.0235$ & 95.9\% \\
Top 6 (SHAP)              &  6 &  5 & 0.9787 & 0.9001 & \textbf{0.9503} & $+0.0502$ & 97.1\% \\
Top 6 (3-method)          &  6 &  5 & 0.9756 & 0.9048 & \textbf{0.9228} & $+0.0180$ & 94.6\% \\
\hline
\end{tabular}
}
\end{table}

The three classes follow different colour--redshift tracks (Fig.~\ref{fig:colourz};
details in Appendix~\ref{app:colour_z}). The with-redshift model has
learned a signal that mixes physics and the SDSS targeting: stars are
identified physically by $z\approx0$, and the apparent size is a
physical discriminant, but the near-absence of SDSS galaxies above
$z\approx1$ lets the model use $z\gtrsim1.5$ as a quasar proxy. Its
GALAXY recall is $0.703$ at $1\le z<1.5$ and $0.016$ at $1.5\le z<2$,
whereas the model without redshift reaches $0.820$ and $0.680$ in the
same bins; the \num{306} galaxies at $1.5\le z<2$ in the SDSS test set
are only $0.08\%$ of the test galaxies ($0.06\%$ of the test set), so
the $0.016$ recall has a small effect on the macro-$F_1$, but in a
Rubin-like deep survey the galaxy distribution extends well beyond
$z=1.5$ and the same failure would affect a substantial fraction of the
sample; on DESI it already does (Sect.~\ref{sec:desi}). The with-redshift
model is therefore not directly applicable to a deep photometric survey,
because its redshift dependence encodes the SDSS target selection. The
price of the no-redshift model is its sensitivity to the magnitude
range (Sect.~\ref{sec:shift}) and its low quasar recall for $z\ge4$;
the separation of quasar light from host-galaxy light remains an active
observational challenge even with JWST \citep{ding2023}.

\begin{figure*}[htbp]
\centering
\includegraphics[width=0.7\textwidth]{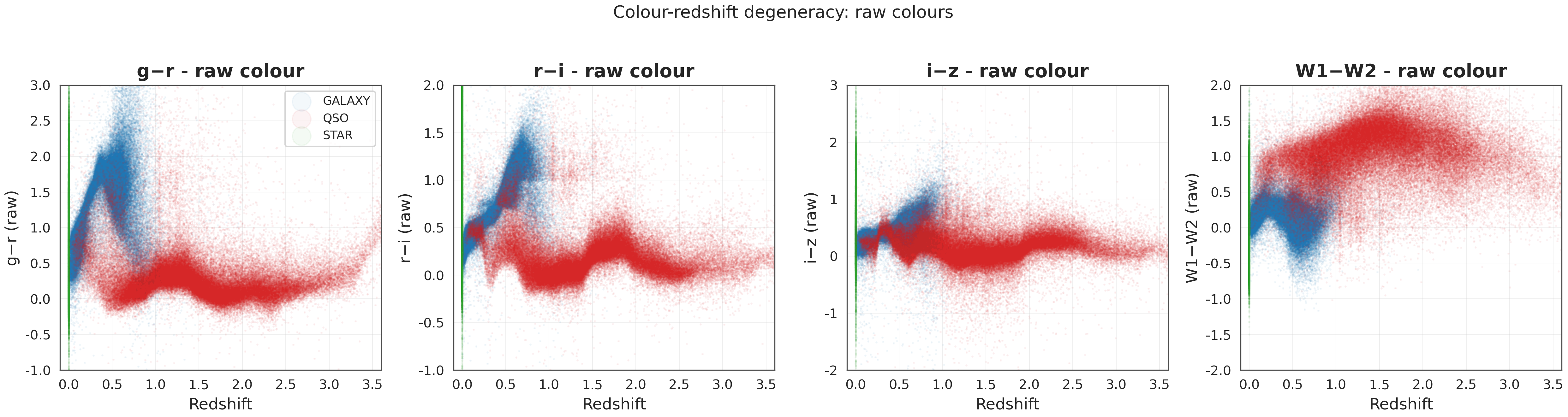}
\caption{Raw colours versus redshift.}
\label{fig:colourz}
\end{figure*}

\paragraph{Internal photo-$z$.}
We build a photo-$z$ feature from the fourteen photometric features (no
raw magnitudes) using gradient-boosted trees
(\texttt{HistGradientBoostingRegressor}, 150 iterations, depth 10) and
a $k$-nearest-neighbour regressor ($k=20$, inverse-distance weights, on
a $1.5\times10^4$-source subsample per fold), with five stratified folds
and out-of-fold predictions. The HGBR photo-$z$ reaches
$\sigma_{\rm NMAD}=0.034$ ($8.2\%$ at $|\Delta z|/(1+z)>0.15$) and the
$k$NN photo-$z$ $\sigma_{\rm NMAD}=0.033$ ($9.6\%$ at the same
threshold), where
$\sigma_{\rm NMAD}=1.48\times{\rm median}(|\Delta z-{\rm median}(\Delta
z)|/(1+z))$. The HGBR variant changes the macro-$F_1$ of the
14-feature model from $0.9561$ (no $z$) to $0.9562$ ($+0.0001$), and
the $k$NN variant to $0.9560$ ($-0.0001$). Both are indistinguishable
from the baseline: because the photo-$z$ is a deterministic function of
the classifier's own features, it is necessarily redundant. This is not
a test of a state-of-the-art photo-$z$; it shows that the colours
already encode much of the redshift information. A state-of-the-art
photo-$z$ that uses template fitting or additional bands might add
independent information; we did not test this.

\paragraph{External photo-$z$.}
To test an independent photo-$z$, we use the neural-network catalog of
\citet{tian2026}, trained on DESI DR1 and SDSS DR19 spectroscopy and
applied to LSDR10 and PS1DR2 imaging. The catalog provides, for each
source, the expected value and standard deviation of the redshift PDF;
it covers the range $0.001<z<2$ and is optimised for galaxies. Of the
\num{1764390} galaxies in the training set and \num{378084} in the test
set, $97.4\%$ have a catalog counterpart within $1''$. Only $1$ of the
\num{792227} non-galaxy sources in the training set (quasars and stars
together) has a catalog counterpart, and none in the test set: the
catalog contains essentially no quasars or stars.

A naive use of the catalog, with a sentinel value for sources without a
match, gives a spuriously high macro-$F_1$ of $0.986$, because the
availability of a photo-$z$ is itself a near-perfect galaxy detector.
We therefore impute every missing value (galaxies without a match, all
quasars, all stars) with a random draw from the matched-galaxy
photo-$z$ distribution, so that the photo-$z$ column cannot act as a
class detector. With this random-draw imputation, the external
photo-$z$ changes the macro-$F_1$ from $0.9561$ (14 photometric
features) to $\mathbf{0.9620}$ (mean only) and $0.9652$ (mean and
standard deviation), a gain of $+0.0059$ and $+0.0091$ respectively.
A control with the median of the matched galaxies in place of the
random draw gives $0.9716$ ($+0.0155$); the difference, $+0.0096$ (from the
unrounded values), is the availability signal, which we cannot report as a
genuine improvement. Of the total $+0.0155$ in the median-impute variant,
$38\%$ comes from the photo-$z$ value and $62\%$ from the availability
of a photo-$z$.

The result is a cautionary example of a selection effect that is easy
to overlook. An external photo-$z$ adds independent information about
galaxy redshifts, but a classifier that relies on the availability of
a photo-$z$ for class discrimination inherits the same prior and
selection dependence as the colour-based classifier
(Sects.~\ref{sec:purity} and~\ref{sec:desi}). We therefore report the
$+0.0059$ (mean) and $+0.0091$ (mean and std) as the genuine
improvement attributable to the external photo-$z$ value, and we
present the $+0.0155$ as an upper bound that includes the selection
effect. Even the corrected gains are modest and are conditional on a
galaxy-only catalog that covers $z<2$; a photo-$z$ catalog that also
covers quasars and stars, or one that extends to higher redshift, might
add more, but we did not test this. An observed improvement in
macro-$F_1$ does not by itself establish that the photo-$z$ adds
independent information: the availability pattern of the catalog is
itself class-dependent, and even the random-draw control could inherit
this dependence at second order. We test this directly by repeating
the random-draw imputation with five independent seeds (42--46): the
macro-$F_1$ varies over the narrow range $0.9621$--$0.9623$, a spread
of $0.0002$ that is well below the bootstrap half-width
($\approx0.0005$). The second-order dependence on the imputation seed
is therefore negligible in this sample; a stronger control that
preserves the per-class abundance of the imputed photo-$z$ would
further tighten this bound, and we have not implemented it. The
positive contribution of the standard deviation ($+0.0032$) shows that
the photo-$z$ uncertainty is itself informative: it allows the
classifier to distinguish galaxies with a sharp photo-$z$ (secure
redshifts) from those with a broad one (ambiguous redshifts), which is
consistent with the well-calibrated PDFs of the catalog.

\begin{table}[htbp]
\caption{Classifier performance with internal and external photo-$z$. The
14-feature baseline is the set of 11 colours plus 3 morphology features
with no raw magnitudes and no redshift (same configuration as the
``Photometric'' row of Table~\ref{tab:optical_only}). The internal rows
use cross-fitted features; the external rows use the catalog of
\citet{tian2026}. The median imputation is dominated by a selection effect,
and the random-draw imputations are the leak-free controls.}
\label{tab:photoz_classification_crossfit}
\centering
\resizebox{\columnwidth}{!}{%
\begin{tabular}{lrrrrr}
\hline\hline
Variant & $n$ feat & macro-$F_1$ & $F_1$ (GAL) & $F_1$ (QSO) & $F_1$ (STAR) \\
\hline
Photometric (14 feat, no $z$)                & 14 & 0.9561 & 0.9799 & 0.9295 & 0.9589 \\
Internal photo-$z$ (HGBR, 14 feat)           & 15 & 0.9562 & 0.9800 & 0.9299 & 0.9588 \\
Internal photo-$z$ ($k$NN, 14 feat)          & 15 & 0.9560 & 0.9799 & 0.9296 & 0.9585 \\
\hline
External photo-$z$, median impute            & 15 & 0.9716 & 0.9906 & 0.9556 & 0.9687 \\
External photo-$z$, random-draw (mean)       & 15 & \textbf{0.9620} & 0.9839 & 0.9398 & 0.9623 \\
External photo-$z$, random-draw (mean + std) & 16 & \textbf{0.9652} & 0.9860 & 0.9462 & 0.9633 \\
\hline
\end{tabular}
}
\end{table}

We also train the model on the low-redshift objects of the training set
($z<0.5$) and test it on the high-redshift objects ($z\ge0.5$), and
vice versa (Table~\ref{tab:redshift_transfer} in
Appendix~\ref{app:supplementary}). The QSO $F_1$ is $0.8187$ when
trained at low and tested at high redshift, and $0.4243$ in the reverse
direction; we interpret this test as a measure of the sensitivity to the
redshift range of the training sample (Appendix~\ref{app:colour_z}).

\subsection{Magnitude shift}
\label{sec:shift}

The previous subsection has shown that training and testing at different
redshifts changes the class composition, so we instead use a
magnitude-shift split ($r<20$ versus $r\ge20$), which changes the
apparent brightness and the photometric noise but leaves every class
present on both sides of the split. With redshift, the macro-$F_1$
changes from $0.984$ (same regime) to $0.975$ (across regimes), a drop
of $0.010$ ($1.0\%$ in relative terms); without redshift, it changes
from $0.933$ to $0.851$, a drop of $0.082$ ($8.8\%$ in relative terms;
Table~\ref{tab:cross_z} in Appendix~\ref{app:supplementary}). The
direction of the transfer is strongly asymmetric. Without redshift,
training on $r<20$ and testing on $r\ge20$ gives $0.774$, whereas
training on $r\ge20$ and testing on $r<20$ gives $0.928$ (in-regime
values $0.981$ and $0.885$); with redshift, the two cross values are
$0.964$ and $0.985$ (in-regime $0.992$ and $0.977$).

Three effects probably contribute. The photometric noise is larger for
faint sources, and a model trained on bright sources has not seen noisy
colours. The faint sample is at higher redshift, where the colours of
galaxies are redder (Sect.~\ref{sec:redshift}), so that a colour that
works as a proxy for a nearby bright source fails for a distant faint
one. The faint sample covers a wider range of colours and noise, so that
training on it interpolates when testing on bright sources, whereas the
reverse extrapolates. The redshift, which does not depend on the
apparent brightness, shields the model from the first two effects. The
class mix of the two magnitude ranges is not the same, because the SDSS
spectroscopic targeting reaches different limiting magnitudes, so part
of the drop is a change of the class composition and not a pure loss of
classification power. If a no-redshift model is used on a deep survey,
the training sample should cover the faint magnitudes.

\subsection{Class priors and calibration}
\label{sec:purity}

The macro-$F_1$ is computed with the class fractions of the test set
($\sim69/16/15\%$), which differ from the fractions in a deep imaging
survey. For a fixed ROC curve, the purity (precision) of class $c$
follows from Bayes' theorem,
\begin{equation}
P_c=\frac{\mathrm{TPR}_c\,\pi_c}{\mathrm{TPR}_c\,\pi_c+\mathrm{FPR}_c\,(1-\pi_c)},
\label{eq:purity}
\end{equation}
where $\pi_c$ is the fraction of class $c$ so that a small
contamination from an abundant class matters much more for a rare class.
We report the purity at a completeness (recall) of $0.90$ under three
class mixes (Table~\ref{tab:working_point}; the corresponding
purity--completeness curves are in Fig.~\ref{fig:pc_curves} in
Appendix~\ref{app:supplementary}); we use purity and completeness
interchangeably with precision and recall. The SDSS-like mix is a
reference; the Rubin-like and Roman-like mixes are assumptions of this
work, chosen to illustrate the effect of a lower quasar fraction, and
not measured sky fractions. Here ``SDSS-like'' is the prior of the raw
SDSS catalog ($60.6/18.7/20.7\%$; Sect.~\ref{sec:data}), not the prior
of the WISE-matched test set ($69.0/15.7/15.3\%$); at the test prior
the QSO purity at $0.90$ completeness is $0.994$, so the difference from
the quoted $0.995$ is negligible for the argument. The purity of
galaxies and stars at a completeness of $0.90$ is $0.996$--$0.999$ and
$1.000$ for all three mixes, since their ROC curves are almost ideal.
The quasar purity falls from $0.995$ (SDSS-like) to $0.977$ ($75/5/20$)
and $0.962$ ($80/3/17$): the macro-$F_1$ at the training prior is
therefore a poor guide when the class mix differs substantially from the
training prior, and users should choose the operating point from the
purity--completeness curve of the class of interest and for their prior.
The quasar curves start at a completeness of about $0.89$ and fall
steeply above about $0.96$.

The reason for the fall of the purity is the arithmetic of
Eq.~(\ref{eq:purity}): at a given completeness the false-positive rate
is fixed ($0.0011$ for the QSO class), so the number of contaminants is
proportional to the number of non-quasars, whereas the number of true
quasars is proportional to $\pi_{\rm QSO}$. For an illustrative
catalog of $10^9$ sources at a completeness of $0.90$, the number of
contaminants is about $10^6$ in all three mixes and the purity falls
because the quasars become rarer; if calibrated posteriors $p(c|x)$ are
available, a change of prior can be handled by
$p'(c|x)\propto(\pi'_c/\pi_c)\,p(c|x)$, but this assumes that $p(x|c)$
does not change, which the magnitude-shift test of
Sect.~\ref{sec:shift} and the external test of Sect.~\ref{sec:desi}
show to be incorrect in general. The purity values in this subsection
use the ROC curve of the tuned Random Forest on the Top-12 4-method
unified set, which is the set used for the purity--completeness curves
of Fig.~\ref{fig:pc_curves}; the Top-12 SHAP set gives very similar
purities (at the SDSS-like prior, the QSO purity at $0.90$ completeness
is $0.994$ for the SHAP set, within $0.001$ of the $0.995$ quoted for
the 4-method unified set).

\begin{table}[htbp]
\caption{Purity at a completeness of $0.90$ under the three class mixes,
with uncalibrated and calibrated scores.}
\label{tab:working_point}
\centering
\resizebox{\columnwidth}{!}{%
\begin{tabular}{llrrrr}
\hline\hline
Mix & Fractions & GAL purity & QSO purity & STAR purity & QSO FPR \\
\hline
SDSS-like  & 60/19/21 & 0.996 & \textbf{0.995} & 1.000 & 0.0011 \\
Rubin-like (assumed) & 75/5/20  & 0.998 & \textbf{0.977} & 1.000 & 0.0011 \\
Roman-like (assumed) & 80/3/17  & 0.999 & \textbf{0.962} & 1.000 & 0.0011 \\
\hline
\multicolumn{6}{l}{\footnotesize Isotonic calibration changes all purities by at most $0.0001$.}
\end{tabular}
}
\end{table}

The purity analysis applies Bayes' theorem to a per-class ROC curve. The
result depends on the ROC curve only and not on the absolute calibration
of the probabilities, because the ROC curve is invariant under any
monotonic transformation of the score. We verify this directly with
one-vs-rest isotonic regression \citep{zadrozny2002} fitted on the
validation set and evaluated on the test set. The expected calibration
error (ECE, the bin-weighted absolute difference between mean confidence
and mean accuracy in fifteen equal bins) drops from $0.0052$ to
$0.0004$: the tuned Random Forest is already close to calibrated on
these data, and the calibration step refines a score that is already
nearly calibrated. The purities at a completeness of $0.90$ change by at
most $0.0001$ (Table~\ref{tab:working_point}), as expected for a
monotonic transformation. We use the uncalibrated scores everywhere else
in the paper.

\subsection{Photometric-only deployment}
\label{sec:deployability}

Neither the absolute magnitudes nor the WISE colours will be available
at deployment time on a photometric survey: absolute magnitudes require
a known distance (i.e., a redshift), and WISE is much shallower than
LSST or Roman and will not have a counterpart for most faint sources. We
therefore test three deployment scenarios (Table~\ref{tab:optical_only}):
\emph{photometric} (11 colours and 3 morphology features, no absolute
magnitudes, no redshift), \emph{optical only} (the 5 optical colours and
the 3 morphology features, no WISE data), and \emph{colours only} (all
11 colours, no morphology).

The photometric scenario reaches $0.9561$, essentially equal to the
no-redshift scenario with magnitudes ($0.9550$), so the magnitudes add
no measurable gain in this scenario. Without any WISE data the
classifier reaches $0.9234$, retaining $93.9\%$ of the
with-redshift score; the drop is $3.3$\,pp relative to the photometric
scenario and $6.0$\,pp relative to the with-redshift reference. The
colours-only scenario reaches $0.9448$ ($96.1\%$), which is better than
the optical-only scenario even though it contains no morphology, because
the WISE colours are informative for the GALAXY--QSO separation; the
morphology adds $0.0113$ to the colours. The optical-only score of
$0.9234$ is the appropriate number to quote for a Rubin/LSST catalog
without WISE counterparts. The deployment scores thus bracket the
possible performance: the optical-only scenario is the pessimistic lower
bound, the photo-$z$ scenario the benefit of a colour-derived photo-$z$,
and the spec-$z$ scenario an upper bound that will not be achievable on
a photometric survey.

\begin{table}[htbp]
\caption{Photometric-only deployment scenarios. Retention is relative to the Top-12 (SHAP, with $z$) reference.}
\label{tab:optical_only}
\centering
\resizebox{\columnwidth}{!}{%
\begin{tabular}{lrrr}
\hline\hline
Scenario & $n$ feat & Test macro-$F_1$ & Retention \\
\hline
Top 12 (SHAP, with $z$) [reference] & 12 & 0.9834 & 100.0\% \\
Top 12 (SHAP, no $z$, with magnitudes) & 11 & 0.9550 & 97.1\% \\
\textbf{Photometric (colours + morphology)} & 14 & \textbf{0.9561} & \textbf{97.2\%} \\
Colours only (no morphology) & 11 & 0.9448 & 96.1\% \\
Optical only (no WISE)       &  8 & 0.9234 & 93.9\% \\
\hline
\end{tabular}
}
\end{table}

\section{External validation on DESI DR1}
\label{sec:desi}

As an independent test of the transfer of the classifier, we apply an
SDSS-trained model to spectroscopically classified sources from the
first data release of the Dark Energy Spectroscopic Instrument (DESI
DR1) \citep{desi2026dr1}. The initial extract contains \num{249839}
sources with a spectroscopic class and a Legacy Survey DR10 counterpart
(drawn from the Value-Added Catalog hosted on the Astro Data
Lab\footnote{\url{https://datalab.noirlab.edu/}} by querying
\texttt{desi\_dr1.zpix} and \texttt{desi\_dr1.photometry} with
\texttt{ZWARN}$=0$ and
\texttt{SPECTYPE}$\in\{$GALAXY, QSO, STAR$\}$ and cross-matching with
the Legacy Surveys DR10 catalog; for the survey overview see
\citealt{dey2019}, and for the DR10 data release itself see the Legacy
Surveys documentation \citep{legacysurvey_dr10}. We use only the $g$,
$r$, $z$, $W1$, and $W2$ bands and the size parameter, all of which
are available in every Legacy Surveys release, so the survey overview
of \citet{dey2019} applies to the photometric quantities used here).
Removing sources with
non-positive flux in any of the five photometric bands or with a
non-positive size leaves \num{230235} sources: \num{226967} galaxies
($98.6\%$), \num{2421} quasars ($1.05\%$), and \num{847} stars ($0.37\%$)
(Table~\ref{tab:desi_sample}); this removal is required because the
Legacy Survey WISE photometry is forced photometry and a non-positive
flux is a non-detection, not a measurement. The class prior is
drastically different from the SDSS training sample, and the median
redshifts are $0.32$ (galaxies), $0.56$ (quasars), and $0.00$ (stars);
the DESI quasar sample is dominated by intermediate-redshift sources
(84th percentile $z=1.43$), whereas the SDSS quasar sample peaks near
$z\approx1.2$, and the DESI class definition may include
lower-luminosity broad-line sources.

The Legacy Survey photometry does not provide the $u$ and $i$ bands, so
we retrained the model on SDSS using only the features that can be built
from $g$, $r$, $z$, $W1$, $W2$, and the size: $g-r$, $r-z$, $g-z$,
$W1-W2$, $z-W1$, $r-W2$, $\log_{10}R$, compactness, and $\mu_r$ (nine
features) plus the redshift (ten features). The Legacy Survey WISE
photometry is in the AB system; we convert to Vega per band
($W1_{\rm Vega}=W1_{\rm AB}-2.699$, $W2_{\rm Vega}=W2_{\rm AB}-3.339$;
\citealt{jarrett2011}). The Legacy Survey \texttt{shape\_r} is a
S\'ersic effective radius, not an SDSS Petrosian radius, and is
systematically smaller (Sect.~\ref{sec:crossmatch}); we do not
recalibrate it here.

On SDSS, the 10-feature model (with $z$) reaches a QSO $F_1$ of
$0.9571$ (recall $0.943$, precision $0.971$), the nine-feature model
without $z$ reaches $0.9207$, and the three optical colours alone reach
$0.764$. On DESI the same models give a QSO recall of $0.469$ with $z$
and $0.547$ without $z$, but a precision of only $0.059$ and $0.051$,
hence a QSO $F_1$ of $0.1046$ and $0.0927$
(Table~\ref{tab:desi_ablation}); no feature subset exceeds $0.105$, and
on SDSS all subsets except the optical-only one stay above $0.90$. The
STAR class is the most robust (recall $1.00$, AUC $0.9997$). On the
full \num{249839}-source extract including the non-detections, the QSO
$F_1$ of the same 10-feature model is $0.085$ if the non-detections are
imputed at $\sim45$ mag; this is a sensitivity check, not a measurement.
The \num{7.8}\% of the extract with non-positive WISE flux are
non-detections in the Legacy Survey forced photometry and cannot be
converted to magnitudes; they are not sources on which the classifier
has been shown to fail, and the present data do not establish whether
the exclusion biases the classifier. We therefore report $0.105$ on the
valid-photometry sample as the primary result.

\begin{table}[htbp]
\caption{QSO metrics of SDSS-trained models on the SDSS test set and on the DESI DR1 sample. The 10-feature model includes redshift; ``no size'' removes $\log_{10}R$, compactness, and $\mu_r$. The DESI QSO $F_1$ is dominated by the $1\%$ quasar prior, and the DESI macro-$F_1$ is not comparable with the SDSS value.}
\label{tab:desi_ablation}
\centering
\resizebox{\columnwidth}{!}{%
\begin{tabular}{llrrrr}
\hline\hline
Survey & Feature set & $n$ feat & QSO $F_1$ & QSO recall & QSO precision \\
\hline
SDSS & 10 feat, with $z$           & 10 & 0.9571 & 0.9432 & 0.9714 \\
SDSS & 9 feat, no $z$              &  9 & 0.9207 & 0.9000 & 0.9424 \\
SDSS & 6 feat, no size             &  6 & 0.9046 & 0.8855 & 0.9245 \\
SDSS & 5 feat, no size $+ W1-W2$   &  5 & 0.9042 & 0.8851 & 0.9242 \\
SDSS & 3 feat, optical only        &  3 & 0.7640 & 0.7272 & 0.8046 \\
\hline
DESI & 10 feat, with $z$           & 10 & \textbf{0.1046} & 0.4688 & 0.0588 \\
DESI & 9 feat, no $z$              &  9 & 0.0927 & 0.5469 & 0.0507 \\
DESI & 6 feat, no size             &  6 & 0.1022 & 0.4688 & 0.0573 \\
DESI & 5 feat, no size $+ W1-W2$   &  5 & 0.1017 & 0.4647 & 0.0571 \\
DESI & 3 feat, optical only        &  3 & 0.0517 & 0.3201 & 0.0281 \\
\hline
\end{tabular}
}
\end{table}

The one-vs-rest ROC curves are in Fig.~\ref{fig:desi_roc} in
Appendix~\ref{app:supplementary}; the AUC values on the \num{230235}
DESI sources are $0.867$ (GALAXY), $0.827$ (QSO), and $0.9997$ (STAR),
against $0.998$, $0.997$, and $1.000$ on the SDSS test set
(Table~\ref{tab:per_class_best}). The QSO AUC drops by $0.17$ and the
GALAXY AUC by $0.13$; the STAR AUC is essentially unchanged. The QSO
purity at $0.90$ completeness is $0.016$ for a $1\%$ prior and $0.076$
for a $5\%$ prior, with a false-positive rate of $0.573$: the
classifier retains substantial ranking power on DESI but has an enormous
false-positive tail, so the precision at any realistic prior is very
low.

The per-class recall in redshift bins (Fig.~\ref{fig:desi_recall_z};
Table~\ref{tab:desi_recall_binned}) is strongly asymmetric. GALAXY
recall is $0.994$ for $z<0.5$ (\num{152710} sources) and $0.944$ for
$0.5\le z<1$ (\num{54806}), drops to $0.288$ for $1\le z<1.5$
(\num{17282}), and is exactly zero for $1.5\le z<2$ (\num{2070} DESI
galaxies, all classified as quasars); QSO recall is $0.308$ for
$z<0.5$ (\num{1112}), $0.375$ for $0.5\le z<1$ (\num{800}), $0.867$
for $1\le z<1.5$ (\num{128}), and $1.000$ for $z\ge1.5$; STAR recall
is $1.000$ in the only bin where stars are present ($z<0.5$,
\num{626} sources). This is the redshift signal of
Sect.~\ref{sec:redshift} operating at full strength: the model has
learned from the SDSS spectroscopic sample that $z\gtrsim1.5$ means
``quasar'', and assigns that class to all galaxies at those redshifts
in DESI, where 2,070 galaxies lie at $1.5\le z<2$ versus 306 in the
SDSS test set. The DESI test shows that the effect is a systematic
failure, not a small tail of the SDSS sample.

\subsection{What causes the drop? A cross-match with SDSS}
\label{sec:crossmatch}

To separate the photometric system, the label definition, and the
intrinsic difficulty of the DESI population, we cross-match the DESI
sample with SDSS by position ($<1''$) and retain \num{11514} common
objects (galaxy-dominated: \num{11385} GALAXY, 100 QSO, 29 STAR). The
labels agree on $99.3\%$ of the objects, but for quasars only $68/100$
SDSS QSOs are confirmed by DESI and $41$ SDSS galaxies are reclassified
as QSOs, so the two surveys define the quasar class on different
populations; the label definitions disagree on $32\%$ of SDSS QSOs and
$38\%$ of DESI QSOs in the common sample. The per-band offsets are
$\Delta g = -0.12$, $\Delta r = -0.13$, $\Delta z = -0.04$,
$\Delta W1 = -0.38$, $\Delta W2 = -0.40$ mag, and
$\Delta\log_{10}R = -0.22$ dex; the Legacy Survey \texttt{shape\_r} is a
S\'ersic effective radius and is systematically smaller than the SDSS
Petrosian radius. Table~\ref{tab:desi_feature_comparison} shows the medians:
DESI galaxies are $\approx0.5$ mag bluer in $g-r$ and $\approx1.0$ mag bluer
in $g-z$ than SDSS galaxies, DESI quasars are $\approx0.6$ mag redder
in $g-r$ and $\approx1.0$ mag redder in $g-z$, and $\log_{10}R$ is
offset by $-0.51$ dex for galaxies and $-0.37$ dex for quasars. Both
effects push the DESI sample toward the classification boundary.

\begin{table}[htbp]
\caption{Feature medians on the SDSS training set and the DESI sample,
per class. The colour plane is shifted by $\approx0.5$ mag in $g-r$ and
$\approx1.0$ mag in $g-z$, and the size parameter by $\approx0.4$ dex.}
\label{tab:desi_feature_comparison}
\centering
\resizebox{\columnwidth}{!}{%
\begin{tabular}{llrrr}
\hline\hline
Class & Feature & SDSS median & DESI median & Offset \\
\hline
GALAXY & $g-r$        & 1.454 & 0.994 & $-0.46$ \\
GALAXY & $r-z$        & 1.118 & 0.736 & $-0.38$ \\
GALAXY & $g-z$        & 2.690 & 1.716 & $-0.97$ \\
GALAXY & $W1-W2$      & 0.171 & 0.203 & $+0.03$ \\
GALAXY & $\log_{10}R$ & 0.536 & 0.025 & $-0.51$ \\
GALAXY & $\mu_r$      & 23.45 & 21.51 & $-1.94$ \\
\hline
QSO & $g-r$        & 0.233 & 0.811 & $+0.58$ \\
QSO & $g-z$        & 0.512 & 1.519 & $+1.01$ \\
QSO & $W1-W2$      & 1.079 & 0.669 & $-0.41$ \\
QSO & $\log_{10}R$ & 0.129 & $-0.241$ & $-0.37$ \\
QSO & $\mu_r$      & 22.25 & 20.54 & $-1.71$ \\
\hline
STAR & $g-r$       & 0.632 & 1.294 & $+0.66$ \\
STAR & $g-z$       & 1.020 & 2.572 & $+1.55$ \\
\hline
\end{tabular}
}
\end{table}

The most striking difference is the WISE colours. On the common objects
(98.9\% galaxies) the Pearson correlation between the two surveys is
$0.81$ for $g-r$, $0.65$ for $r-z$, and $0.87$ for $g-z$, but only
$0.004$ for $W1-W2$ and $-0.005$ for $r-W2$. This near-zero Pearson
value is misleading: the Spearman correlation of $W1-W2$ is $0.72$
(95\% CI $0.71$--$0.73$), the $3\sigma$-trimmed Pearson is $0.88$, and
the robust scatter of the $W1-W2$ difference is $0.057$ mag
(Table~\ref{tab:bootstrap_three_models} and
Table~\ref{tab:w1_bins}). The two surveys therefore agree on the
$W1-W2$ colour of a typical object at the $0.06$ mag level; the
Pearson correlation is driven to zero by heavy tails and by the narrow
dynamic range of the galaxy-dominated sample, not by a genuine absence
of correlation. The agreement nevertheless degrades with depth: in
quantile bins of SDSS $W1$ the Spearman correlation of $W1-W2$ peaks
at $0.86$ in $13.0\le W1<13.5$ and falls to $0.50$ in the faintest
common bin ($14.3\le W1<17.8$), and the robust scatter of the
difference rises from $0.038$ to $0.087$ mag over the same range
(Table~\ref{tab:w1_bins}). The WISE colours therefore do not transfer
between the two photometric systems at the depth of the DESI sample,
but the loss is a depth effect, not a class-wise decorrelation.

We evaluate the SDSS-trained models on the common objects under four
conditions: SDSS photometry with SDSS labels (A), DESI photometry with
SDSS labels (B), DESI photometry with DESI labels (C), and DESI
photometry after a cross-fitted linear recalibration to SDSS (D). For
the 10-feature model the QSO $F_1$ changes from $A=0.449$ to $B=0.395$
(photometric effect $-0.055$), from $B$ to $C=0.333$ (label effect
$-0.061$), and from $B$ to $D=0.309$ (calibration effect $-0.086$);
the QSO AUC drops from $0.957$ (A) to $0.908$ (B) to $0.886$ (C) and is
recovered to $0.929$ (D). To quantify the
uncertainty we resample the 11,514 common objects
jointly for all conditions (paired bootstrap, 2000 resamples;
Table~\ref{tab:bootstrap_three_models}). On the QSO AUC the photometric
effect is resolved at $-0.049$ (95\% CI $-0.093$ to $-0.013$), whereas
the label effect is not ($-0.022$, 95\% CI $-0.070$ to $+0.023$) and
the contrast $|$label$|-|$photometry$|$ is not resolved either (95\% CI
$-0.081$ to $+0.043$). The decomposition therefore establishes the
photometric-system effect on AUC but not the relative size of the label
effect. On the QSO $F_1$ the photometric effect is not resolved
($-0.055$, 95\% CI $-0.148$ to $+0.037$), while both the label effect
($-0.061$, 95\% CI $-0.122$ to $-0.002$) and the calibration effect
($-0.086$, 95\% CI $-0.159$ to $-0.019$) are resolved. The
recalibration thus improves the QSO AUC (by $+0.020$, 95\% CI $-0.0002$
to $+0.041$, two-sided bootstrap tail probability $p=0.053$) but degrades
the QSO $F_1$; the exact mechanism for this divergence is not explored in
this work. These results are obtained on a
sample that is $98.9\%$ galaxies and contains only $100$ common QSOs;
they are a controlled decomposition on the same physical objects, not a
measurement of the DESI performance. The rest of the drop from the SDSS
test ($0.957$) to A ($0.449$) is the prior.

Table~\ref{tab:bootstrap_three_models} repeats the paired bootstrap for
the 9-feature (no-$z$) and 6-feature (no-size) models on the same
\num{11514} common objects. The photometric-system effect on the QSO
AUC is resolved at 95\% only for the 10-feature model ($-0.049$, 95\%
CI $-0.093$ to $-0.013$); for the 9-feature model ($-0.025$, 95\% CI
$-0.057$ to $+0.002$) and the 6-feature model ($-0.014$, 95\% CI
$-0.052$ to $+0.023$) the point estimates are negative but the
intervals include zero, consistent with the lower sensitivity of these
feature sets to the photometric system.

\begin{table}[htbp]
\caption{Paired bootstrap on the \num{11514} common SDSS--DESI
objects for the QSO AUC, photometric effect (B$-$A), for three
feature sets. The resampling is joint across all conditions, so
the object-to-object noise that cancels in a paired comparison
does not inflate the intervals. Only the 10-feature model
resolves the photometric-system effect at 95\% confidence; the
point estimates for the 9-feature (no-$z$) and 6-feature
(no-size) models are negative but the intervals include zero.
With $N=100$ common QSOs, a non-resolved effect is a statement
about statistical power, not about the absence of an effect.}
\label{tab:bootstrap_three_models}
\centering
\resizebox{\columnwidth}{!}{%
\begin{tabular}{lrrrc}
\hline\hline
Model & $\Delta$AUC (B$-$A) & 95\% CI & $p$ & resolved at 95\% \\
\hline
10 features, with $z$ & $-0.049$ & $[-0.093, -0.013]$ & $0.004$ & yes \\
9 features, no $z$ & $-0.025$ & $[-0.057, +0.002]$ & $0.080$ & no \\
6 features, no size & $-0.014$ & $[-0.052, +0.023]$ & $0.478$ & no \\
\hline
\end{tabular}
}
\end{table}

\subsection{WISE depth and the failure of the classifier}
\label{sec:wise}

On the \num{218721} DESI sources not in the common sample, the QSO
$F_1$ is $0.104$ (AUC $0.824$, FPR $0.092$), consistent with the
$0.105$ on the full sample. Restricting to sources whose $W1$ and $W2$
magnitudes are within the SDSS 99th-percentile envelope ($W1\le17.6$,
$W2\le17.0$) raises the QSO $F_1$ to $0.288$ (recall $0.475$,
precision $0.207$, FPR $0.024$, AUC $0.867$), a factor of $2.8$ in
$F_1$ and $3.8$ in $1/{\rm FPR}$. The SDSS WISE catalog has a median
$W1$ of $15.3$ mag and a 99th percentile of $17.6$, whereas the Legacy
Survey forced photometry used for DESI reaches a 99th percentile of
$20.4$ mag (Table~\ref{tab:wise_depth}); beyond the 99th percentile
of the SDSS $W1$ distribution ($W1>17.6$), the WISE colours become
increasingly noise-dominated and the classifier over-predicts quasars.
This depth dependence is visible directly in the common
objects: in quantile bins of SDSS $W1$ the Spearman correlation of
$W1-W2$ peaks at $0.86$ in $13.0\le W1<13.5$ and falls to $0.50$ in
the faintest common bin ($14.3\le W1<17.8$), and the robust scatter of
the difference rises from $0.038$ to $0.087$ mag over the same range
(Table~\ref{tab:w1_bins}). A linear recalibration learned on
the common objects does not help ($F_1=0.099$ on the full DESI-only
sample and $0.262$ on the restricted sample), consistent with the
non-linearity of the photometric offsets. A major transfer failure,
and in this sample the largest single one, is therefore the WISE
photometry of faint sources; this test is confounded with optical depth
and redshift, because bright WISE sources are also brighter optically,
so it does not establish that the classifier itself is not deficient.
A matched-depth training sample or a WISE selection cut at the
99th-percentile envelope of the SDSS $W1$ distribution are the simplest
remedies, and a retraining on the target photometric system is likely
necessary for a Rubin- or Roman-like survey.

\begin{table}[htbp]
\caption{WISE depth of the SDSS training sample, the DESI sample, and
the common objects. The DESI sample extends $\approx3$ mag deeper in
$W1$ and $W2$ at the 99th percentile. About $7.8\%$ of the sources in
the raw DESI extract have non-positive WISE flux (non-detections in the
Legacy Survey forced photometry) and are excluded from the valid
sample. $p99$ is the 99th percentile of the valid sample, not a
detection limit; the faintest common object is at $W1\approx17.8$.}
\label{tab:wise_depth}
\centering
\resizebox{\columnwidth}{!}{%
\begin{tabular}{lrrrrr}
\hline\hline
Sample & $N$ & $W1_{p50}$ & $W1_{p99}$ & $W2_{p50}$ & $W2_{p99}$ \\
\hline
SDSS (valid)   & 3,652,310 & 15.3 & 17.6 & 14.9 & 16.6 \\
DESI (valid)   &   230,235 & 16.1 & 20.4 & 15.6 & 19.5 \\
Common objects &    11,514 & 15.2 & 17.8 & 14.8 & 16.9 \\
\hline
\end{tabular}
}
\end{table}

\begin{table}[htbp]
\caption{Agreement of $W1-W2$ between SDSS and DESI on the
\num{11514} common objects, in bins of the SDSS $W1$ magnitude.
$\rho_s$ is the Spearman correlation with a bootstrap 95\%
interval; $\sigma_{\rm rob}$ is the robust scatter
($1.4826\times$MAD) of the DESI$-$SDSS colour difference, and
$f_{|\Delta|>1}$ is the fraction of objects with
$|\Delta(W1-W2)|>1$ mag. The correlation peaks at $0.86$ in
$13.0\le W1<13.5$ and falls to $0.50$ in the faintest bin
($14.3\le W1<17.8$); the robust scatter rises from $0.038$ to
$0.087$ mag over the same range, which is the depth effect
discussed in Sect.~\ref{sec:wise}.}
\label{tab:w1_bins}
\centering
\resizebox{\columnwidth}{!}{%
\begin{tabular}{lrrrrr}
\hline\hline
$W1$ bin & $N$ & $\rho_s(W1-W2)$ & 95\% CI &
$\sigma_{\rm rob}(W1-W2)$ & $f_{|\Delta|>1}$ \\
\hline
$9.5\le W1<13.0$ & 2300 & $0.843$ & $[0.826, 0.860]$ & $0.038$ & $0.09\%$ \\
$13.0\le W1<13.5$ & 2304 & $0.862$ & $[0.845, 0.876]$ & $0.033$ & $0.00\%$ \\
$13.5\le W1<13.9$ & 2304 & $0.806$ & $[0.783, 0.824]$ & $0.042$ & $0.09\%$ \\
$13.9\le W1<14.3$ & 2301 & $0.723$ & $[0.695, 0.748]$ & $0.051$ & $0.13\%$ \\
$14.3\le W1<17.8$ & 2305 & $0.496$ & $[0.461, 0.533]$ & $0.087$ & $0.39\%$ \\
\hline
\end{tabular}
}
\end{table}
In summary, the SDSS-trained model does not transfer to DESI without
adaptation: its star classification remains essentially perfect
(recall $1.00$, AUC $0.9997$, $N=\num{847}$ stars), but the quasar
precision collapses. The cross-match of the \num{11514} common objects
and the WISE depth analysis decompose the loss into three contributions;
the largest recovery in DESI QSO $F_1$ is obtained by restricting to the
SDSS WISE range. First, the photometric
system changes the QSO AUC by a point estimate of $-0.05$,
resolved at the 95\% level for the 10-feature model, and the agreement
of the $W1-W2$ colour degrades with depth (Spearman peaks at $0.86$ in
$13.0\le W1<13.5$ and falls to $0.50$ in $14.3\le W1<17.8$). Second,
the label definitions disagree on $\sim32\%$ of SDSS QSOs; the point
estimate of the label effect on the QSO AUC is $-0.02$, but it is not
statistically resolved on the 100 common QSOs. Third, the WISE depth of
the DESI-only sample: restricting to the SDSS WISE range recovers a
factor $2.8$ in $F_1$. A retraining on the target photometric system, a
matched-depth training sample, or a WISE selection cut at the
99th-percentile envelope of the SDSS $W1$ distribution are the natural
next steps.

\section{Discussion and conclusions}
\label{sec:discussion}

\subsection{Discussion}
\label{sec:discussion_main}

With redshift, engineered features yield no measurable gain: the Top-12 set changes the macro-$F_1$ by only $0.0004$ over the 11 raw features ($0.9834$ vs $0.9830$), and the lowest-ranked features add nothing. Without redshift, engineered features add $0.0044$ ($0.9507$ to $0.9551$), roughly ten times the bootstrap half-width ($\approx0.0005$). Group ablation shows this gain comes entirely from colours (optical $+0.0024$, WISE $+0.0013$), while morphology alone hurts ($-0.0015$). Although ten of the Top-12 features are engineered, rankings of correlated features spread credit arbitrarily, and per-feature importance can be near-zero for near-duplicates with sizeable joint importance. The high rank of $W1-W2$ does not establish unique information (removing it changes the macro-$F_1$ by $-0.0002$); rather, the mid-infrared colours as a correlated group carry partly independent information.

Redshift is the most important feature, but its importance is a survey-dependent signal mixing physics (the stellar peak at $z\approx0$) with the SDSS selection function (the near-absence of galaxies above $z\approx1$). Without it, a 12-feature set retains $97.1\%$ of the macro-$F_1$ because colours already encode much of the galaxy redshift. An internal photo-$z$ changes the macro-$F_1$ by only $\pm0.0001$ (necessarily redundant). An external photo-$z$ adds $+0.0059$ to $+0.0091$ after correcting for its galaxy-only selection; a naive median imputation would report $+0.0155$, of which $62\%$ is a selection effect, and the corrected gain is stable against the imputation seed (a spread of $0.0002$ over five seeds). The deployment scores bracket the possible performance: the optical-only scenario ($0.9234$) is the pessimistic lower bound for a catalog without WISE, and the spec-$z$ scenario is an upper bound unreachable on a photometric survey.

Four further properties follow: (1) magnitudes are not a hidden redshift proxy, as the photometric scenario matches the no-redshift scenario with magnitudes ($0.9561$ vs $0.9550$); (2) the class-prior analysis of Sect.~\ref{sec:purity} is the relevant deployment metric, not the training macro-$F_1$; (3) the redshift signal means the with-redshift model is not the better choice for deep surveys, as the DESI test confirms; and (4) a major transfer failure, and in this sample the largest single one, is not the classifier but the WISE photometry of faint sources: the QSO $F_1$ rises from $0.104$ to $0.288$ when restricting to the SDSS WISE range; this test is confounded with optical depth and redshift, so it does not establish that the classifier itself is not deficient.

These results are consistent with class physics. Stars ($z\approx0$, point-like) lose the most $F_1$ when redshift is removed ($0.9987\to0.9518$), as separation must rely on UV excess and mid-IR colours. Galaxies (extended, $4000$\,\AA\ break) allow colours to estimate redshift. Quasars (blue continuum, mid-IR excess) make $W1-W2$ and optical colours crucial; losing WISE bands costs $3.3$\,pp. Weaknesses have physical causes: low-$z$ quasars with dominant host galaxies, and $z\gtrsim3$ quasars with Lyman-$\alpha$ suppressed $u/g$ fluxes.

The scope of these conclusions is bounded by the SDSS--WISE selection function (redshift range, magnitude limits, WISE depth, and class composition). The galaxy distribution ends near $z\approx1$ and the WISE 99th-percentile depth is $W1=17.6$, both shallower than Rubin, Roman, and DESI. Our conclusions are statements about SDSS--WISE classifiers, not photometric classification in general; deeper or wider surveys may exhibit different feature importances and failure modes.

Of the \num{6949} errors, $96.8\%$ are galaxy--quasar confusions where classes overlap in redshift ($z\lesssim1$). Low-$z$ quasars are lost to galaxies (QSO recall $0.28$ at $z<0.05$, due to dominant host-galaxy light), and high-$z$ galaxies are lost to quasars if redshift is available (GALAXY recall $0.016$ at $1.5\le z<2$).

Published per-class scores vary due to different samples, redshift ranges, and metrics. Re-running simplified versions of three published methods on our fixed test set yields agreement within $0.0025$ (with redshift) and $0.0027$ (without) in macro-$F_1$, and within $0.007$ and $0.0049$ in QSO $F_1$. Thus, for tree-based classifiers on this sample, the feature list is a secondary factor; the spread in literature reflects sample, cross-match, and evaluation choices. This is limited to our four reconstructed feature lists and does not extend to other algorithms (e.g., deep learning).

\subsection{Limitations}
\label{sec:limitations}

The SDSS spectroscopic sample is not representative of the photometric population due to its target selection (e.g., the main galaxy sample is limited at $r<17.77$ while quasars reach fainter magnitudes), causing a class composition that varies strongly with magnitude. Requiring a WISE counterpart removes $21.3\%$ of the sample (class-dependent losses of $10.3\%$ for galaxies, $33.7\%$ for quasars, and $42.1\%$ for stars). Because WISE is shallower than LSST or Roman, most faint sources will lack a counterpart (quantified by the optical-only scenario in Sect.~\ref{sec:deployability}). The Rubin- and Roman-like class mixes in Sect.~\ref{sec:purity} remain illustrative assumptions.

Our internal photo-$z$ test only measures redundancy and does not evaluate a state-of-the-art estimator. The external photo-$z$ catalog \citep{tian2026} is galaxy-only ($97.4\%$ of galaxies have counterparts, but virtually no quasars or stars). While the random-draw control isolates the true gain of the external photo-$z$ ($+0.0059$ to $+0.0091$ in macro-$F_1$), these gains are strictly conditional on a galaxy-only catalog limited to $z<2$; a catalog covering quasars and higher redshifts might add more. The second-order dependence of the random-draw control on the imputation seed is negligible: repeating the imputation with five independent seeds changes the macro-$F_1$ by at most $0.0002$, well below the bootstrap half-width ($\approx0.0005$).

The DESI validation uses a reduced feature set, AB-to-Vega conversion, and a different size parameter. The paired bootstrap cannot resolve the label-definition effect because the common sample contains only $100$ QSOs (a limitation of statistical power, not an absence of effect). Recalibration improves QSO AUC but degrades QSO $F_1$. The DESI loss is dominated by WISE depth; this test is confounded with optical depth and redshift, and it does not by itself establish that the classifier is not deficient. Retraining on DESI photometry or applying a WISE selection cut at the 99th-percentile envelope of the SDSS $W1$ distribution are the natural next steps. Finally, the reproductions in Table~\ref{tab:lit_baselines} are simplified versions of the original pipelines, the classifiers are restricted to five non-deep-learning algorithms, and only one external survey is used for validation.

\subsection{Conclusions}
\label{sec:conclusions}

We constructed 14 composite features from SDSS--WISE photometry and
morphology, ranked 25 features with four methods, and tested the value
of the features, of the spectroscopic redshift, of a colour-derived
photometric redshift, and of the evaluation choices with five
classifiers on twelve scenarios, three photometric-only deployment
scenarios, three published methods re-run on a common test set, and an
external sample from DESI DR1. Our main findings are the following.

\begin{enumerate}
\item With spectroscopic redshift, the tuned Random Forest on the Top-12
      (SHAP) set reaches a macro-$F_1$ of $0.9834$ on the full unfiltered
      test set; the per-class scores are $0.991$ (GALAXY), $0.961$ (QSO),
      and $0.999$ (STAR). Removing the redshift costs $3.0$--$3.1$\,pp at
      12 features and $6.4$--$7.9$\,pp at 6 features; re-selection
      recovers $1.8$--$5.0$\,pp of the latter.
\item With redshift, the 13 lowest-ranked features make a practically
      negligible contribution. Without redshift, the full 24-feature set
      gains $0.0044$ over the raw features ($0.9507$ to $0.9551$); the
      gain comes entirely from the colours (optical $+0.0024$, WISE
      $+0.0013$), while the three morphology features add no measurable
      gain once the colours are present and alone degrade the score
      ($-0.0015$). The morphology features are therefore useful only in
      the absence of a continuous size measure, which the raw Petrosian
      radius already provides. The high rank of engineered features
      partly reflects redundancy: near-duplicate morphology features
      vary between resamples (SHAP rank of $\log_{10}R_{\rm petro}$:
      $12.2\pm8.7$), and the morphology cluster, whose members have
      per-feature importances of $0.0001$--$0.0005$, has a joint
      importance of $0.0021$ with redshift and $0.1422$ without it
      ($67.7\times$). The ablation of $W1-W2$ shows a contribution of
      $0.0024$ in the 6-feature set and a negligible contribution in the
      12-feature set.
\item A photo-$z$ built from the classifier's own features changes the
      macro-$F_1$ by $\pm0.0001$ at 14 features; because it is a
      deterministic function of the classifier's features, it is
      necessarily redundant. This is not a test of a state-of-the-art
      photo-$z$. The external photo-$z$ of \citet{tian2026} adds
      $+0.0059$ (mean) and $+0.0091$ (mean and standard deviation) in
      macro-$F_1$ after correcting for the galaxy-only selection of the
      catalog; a naive median imputation would give $+0.0155$, of which
      $62\%$ is a selection effect. The corrected gains are modest and
      conditional on a galaxy-only catalog that covers $z<2$, and they
      are stable against the imputation seed (a spread of $0.0002$ over
      five independent seeds). The
      redshift signal also mixes physics with the SDSS selection
      function: with it, the GALAXY recall is $0.016$ at $1.5\le z<2$
      (against $0.68$ without it), and without it the QSO recall falls
      to about $0.39$ at $z\ge4$ because of the Lyman-$\alpha$ forest.
      The no-redshift model is $8.2\times$ more sensitive to a magnitude
      shift ($-0.082$ versus $-0.010$).
\item The quasar purity at a completeness of $0.90$ falls from $0.995$
      to $0.977$ and $0.962$ for assumed quasar fractions of $5\%$ and
      $3\%$; isotonic calibration changes the purity by at most
      $0.0001$. Tuning a Random Forest is neutral or harmful, while
      XGBoost and LightGBM gain up to $+0.0066$ in the best scenario (mean gains
      up to $+0.0034$). A photometric-only
      scenario reaches $0.9561$, essentially equal to the no-redshift
      scenario with magnitudes ($0.9550$); an optical-only scenario
      reaches $0.9234$ ($93.9\%$ retention).
\item Simplified reproductions of three published methods, each
      evaluated with a default Random Forest on the same test set,
      agree within $0.0025$ (with redshift) and $0.0027$ (without) in
      macro-$F_1$. For tree-based classifiers on this sample, the
      feature list is therefore a secondary factor. The statement does
      not extend to other algorithms (e.g., deep learning on images)
      and does not show that the original pipelines would agree. We
      also note that four of the five classifiers are tree-based and
      none is a deep-learning model, so our results say nothing about
      convolutional or transformer architectures on images.
\item On \num{230235} DESI DR1 sources with valid photometry an
      SDSS-trained model reaches a QSO $F_1$ of $0.105$ (recall $0.47$,
      precision $0.059$, QSO AUC $0.83$), and the GALAXY recall
      collapses to zero for $1.5\le z<2$ (\num{2070} galaxies, all
      classified as quasars). A paired bootstrap on the \num{11514}
      common objects resolves the photometric-system effect on the QSO
      AUC at $-0.05$ (95\% CI $-0.09$ to $-0.01$) for the 10-feature
      model, but the label-definition effect ($-0.02$, 95\% CI $-0.07$
      to $+0.02$) and the contrast between the two effects are not
      statistically resolved on the $100$ common QSOs. A major transfer
      failure, and in this sample the largest single one, is the WISE
      photometry of faint sources: the WISE depth of the DESI sample
      (99th percentile $W1=20.4$) is $\approx3$ mag deeper than the
      SDSS 99th-percentile envelope ($W1=17.6$), and restricting to the
      SDSS-range WISE photometry ($W1\lesssim17.8$) raises the QSO
      $F_1$ from $0.104$ to $0.288$ and reduces the false-positive rate
      from $0.092$ to $0.024$. This test is confounded with optical
      depth and redshift, because bright WISE sources are also brighter
      optically, and it does not establish that the classifier itself
      is not deficient. The near-zero Pearson correlation of $W1-W2$
      ($r=0.004$) on the galaxy-dominated common sample is misleading:
      the Spearman correlation peaks at $0.86$ in $13.0\le W1<13.5$
      and falls to $0.50$ in the faintest common bin
      ($14.3\le W1<17.8$), and the robust scatter of the colour
      difference rises from $0.038$ to $0.087$ mag over the same range
      (Table~\ref{tab:w1_bins}). The transfer failure is therefore
      dominated, in this sample, by the WISE depth of the faint
      sources; this test does not establish that the classifier itself
      is not deficient.
\end{enumerate}

For SDSS--WISE and tree-based classifiers, we therefore recommend
validating feature rankings with raw-only baselines and grouped
importances, reporting metrics on an unfiltered test sample, reporting
purity and completeness for each class under an explicit class prior
and the prior-independent AUC, evaluating a scenario without redshift
and one without WISE, reporting the per-class recall as a function of
redshift, and testing a classifier on an independent survey before
claiming deployability. Our results provide an estimate of the
deployment cost of a Rubin/Roman-like classifier ($93.9\%$ of the
with-redshift performance without WISE) but do not support a claim of
general deployability: they are bounded by the SDSS--WISE selection
function, and the external test shows that a classifier trained on SDSS
requires a correction of the class prior and an adaptation to the
photometric system of the target survey. In summary, our results are
primarily valid for the SDSS--WISE selection function, and transfer to a
deeper survey without recalibration is not guaranteed.

\begin{acknowledgements}
This work was supported by the Iran National Science Foundation (INSF).
This research has made use of
the CasJobs service\footnote{\url{https://skyserver.sdss.org/CasJobs/}},
originally developed by the Johns Hopkins University/Sloan Digital Sky
Survey (JHU/SDSS) team. This research made use of scikit-learn
\citep{scikit-learn}, XGBoost \citep{chen2016}, LightGBM \citep{ke2017},
CatBoost \citep{prokhorenkova2018}, and SHAP \citep{lundberg2017}.
\end{acknowledgements}

\clearpage
\begin{appendix}
\section{Supplementary tables and figures}
\label{app:supplementary}

\begin{figure}[htbp]
\centering
\includegraphics[width=0.45\textwidth]{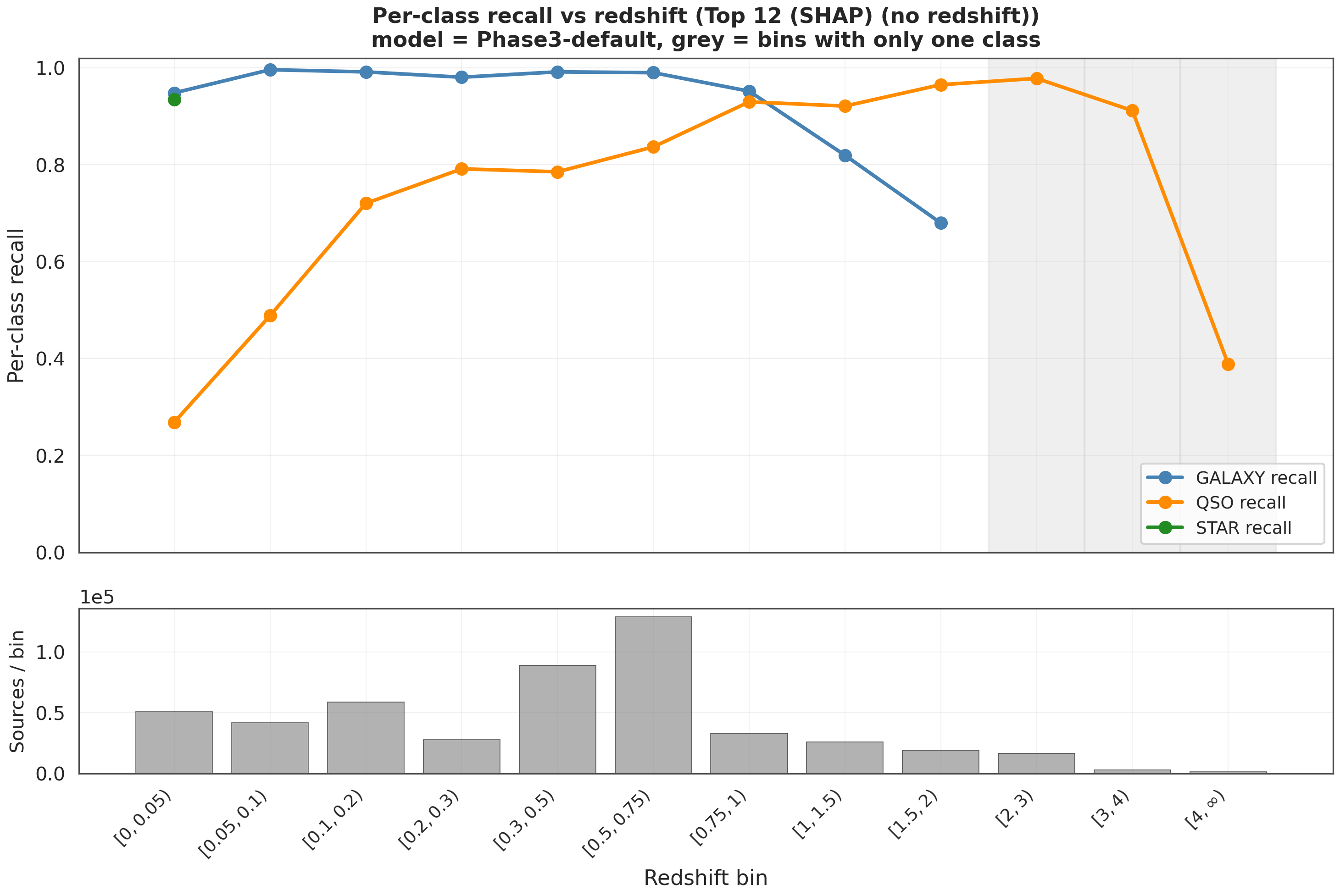}
\caption{As Fig.~\ref{fig:redshift_recall}, but without redshift.}
\label{fig:recall_noz}
\end{figure}

\begin{figure*}[htbp]
\centering
\includegraphics[width=0.7\textwidth]{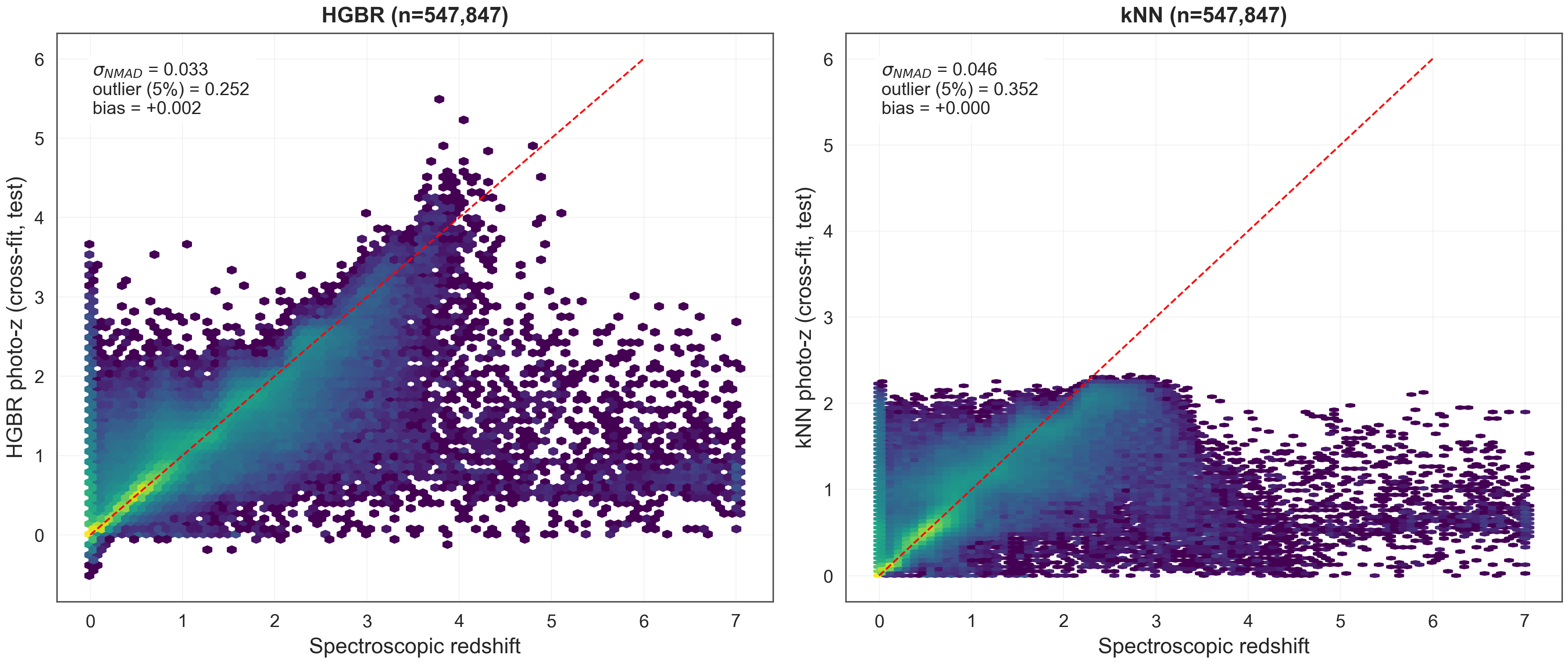}
\caption{Cross-fitted photo-$z$ versus spectroscopic redshift.}
\label{fig:photoz_vs_specz_crossfit}
\end{figure*}

\begin{figure*}[htbp]
\centering
\includegraphics[width=0.7\textwidth]{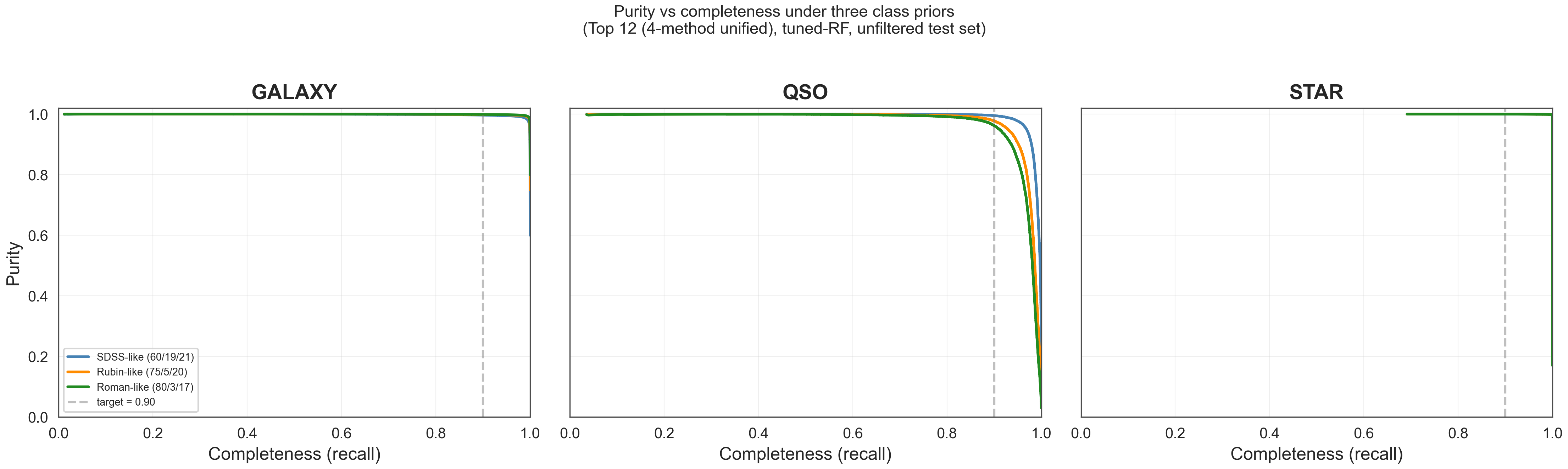}
\caption{Purity--completeness curves for the three class mixes
(SDSS-like $60/19/21$, Rubin-like $75/5/20$, Roman-like $80/3/17$),
computed from the tuned Random Forest on the Top-12 (4-method unified)
set. The dashed line marks a completeness of $0.90$. The GALAXY and
STAR purities stay close to $1$; the quasar curves start at a
completeness of about $0.89$ and fall steeply above about $0.96$.}
\label{fig:pc_curves}
\end{figure*}

\begin{figure}[htbp]
\centering
\includegraphics[width=0.7\columnwidth]{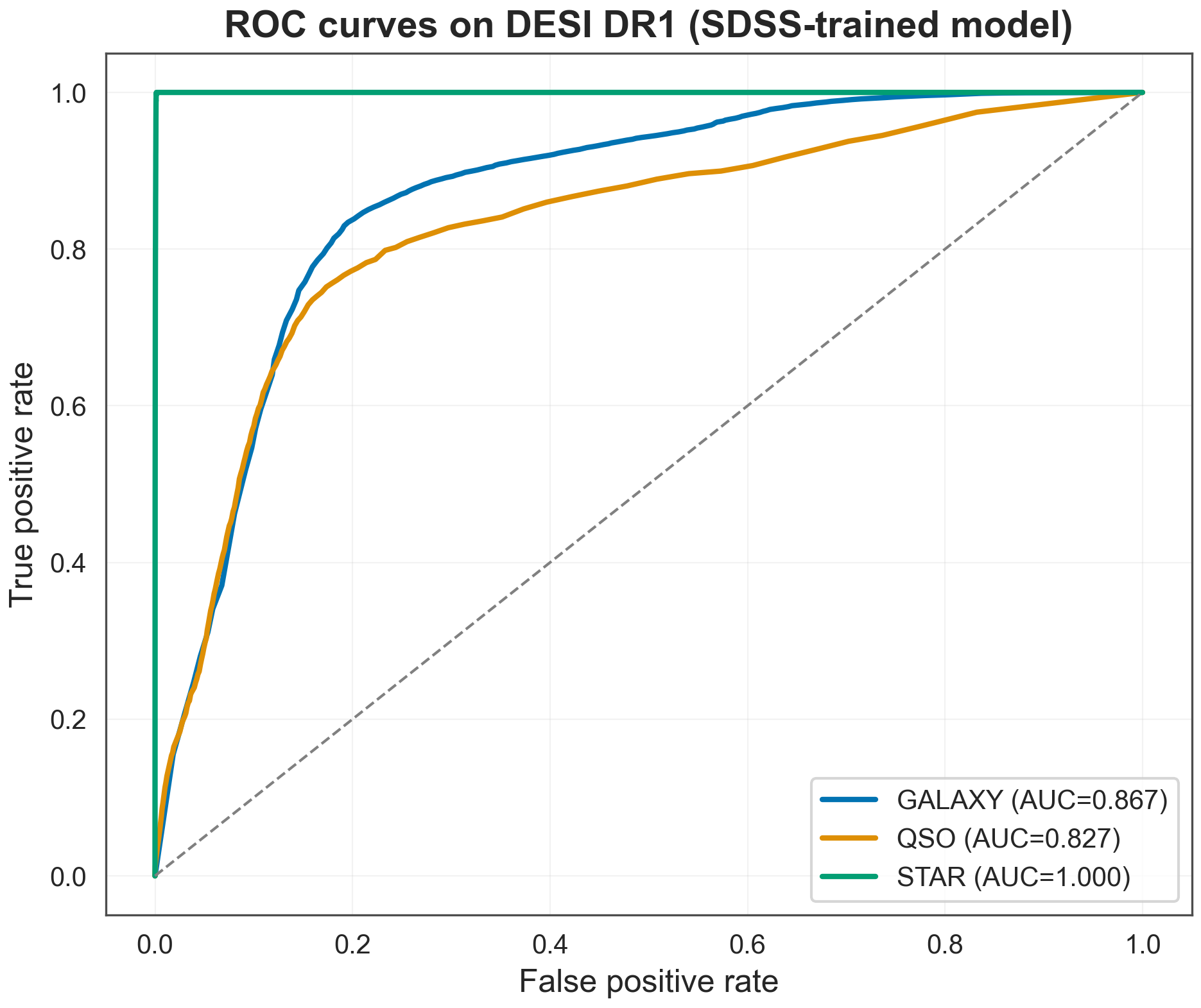}
\caption{Per-class ROC curves on the \num{230235} DESI DR1 sources with
valid photometry, for the 10-feature SDSS-trained model. The AUC values
are $0.867$ (GALAXY), $0.827$ (QSO), and $0.9997$ (STAR); the
corresponding SDSS values of the tuned 12-feature model are $0.998$,
$0.997$, and $1.000$. The QSO AUC drops by $0.17$ and the GALAXY AUC
by $0.13$; the STAR AUC is essentially unchanged.}
\label{fig:desi_roc}
\end{figure}

\begin{figure}[htbp]
\centering
\includegraphics[width=0.7\columnwidth]{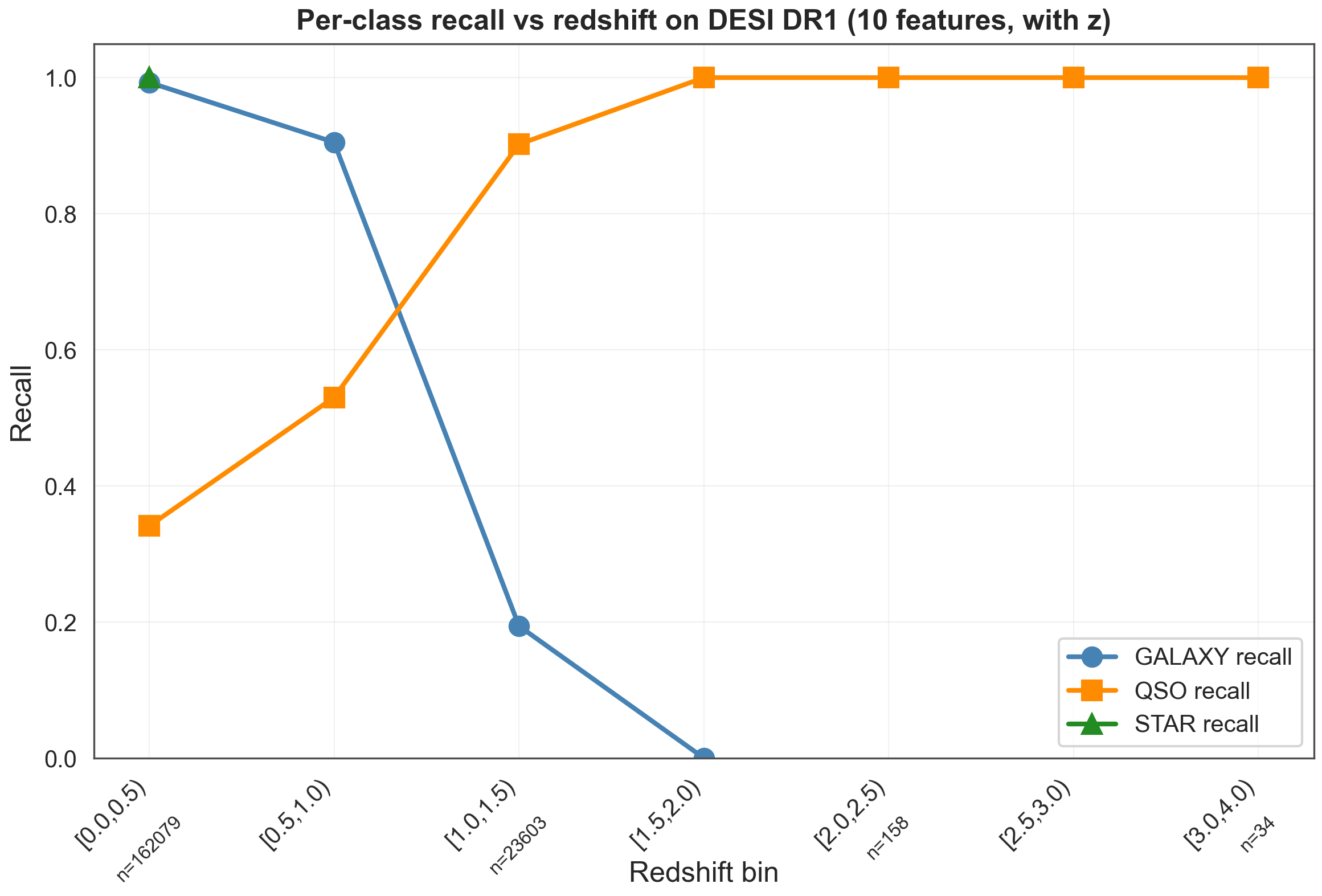}
\caption{Per-class recall as a function of redshift on the \num{230235}
DESI DR1 sources, for the 10-feature SDSS-trained model (with $z$). The
GALAXY recall is $0.994$ for $z<0.5$ and $0.944$ for $0.5\le z<1$,
drops to $0.288$ for $1\le z<1.5$, and is exactly zero for
$1.5\le z<2$ (\num{2070} galaxies, all classified as quasars). The QSO
recall rises from $0.308$ at $z<0.5$ to $1.000$ for $z\ge1.5$. The
number of sources per bin is given at the bottom.}
\label{fig:desi_recall_z}
\end{figure}

This appendix collects the supplementary tables and figures referenced
in the main text. Tables~\ref{tab:bootstrap}--\ref{tab:recall_binned_all}
give the stability of the feature ranking, the per-class $F_1$ of the
tuned scenarios, the full test-set macro-$F_1$, the effect of
hyperparameter tuning, the per-feature permutation importances, the
cross-redshift transfer test, the magnitude-shift test, and the
per-class recall in redshift bins.
Tables~\ref{tab:spearman} and~\ref{tab:lit_baselines} list the
Spearman correlations between redshift and four colours, and the
simplified reproductions of the published methods.
Tables~\ref{tab:desi_sample}--\ref{tab:desi_recall_binned} describe the
DESI DR1 sample, the DESI purity, and the per-class recall in redshift
bins on DESI.
Table~\ref{tab:bootstrap_three_models} reports the three-model paired
bootstrap on the 11,514 common objects, and Table~\ref{tab:w1_bins}
reports the colour diagnostics in bins of SDSS $W1$ used in
Sect.~\ref{sec:crossmatch} and Sect.~\ref{sec:wise}.
Figs.~\ref{fig:recall_noz}--\ref{fig:desi_recall_z} show the per-class
recall without redshift, the normalised mean $|\mathrm{SHAP}|$ per
redshift bin, the cross-fitted photo-$z$, the purity--completeness
curves, the per-class ROC curves on DESI DR1, and the per-class recall
as a function of redshift on DESI DR1.

\subsection{Colour--redshift tracks, SHAP bins, and redshift transfer}
\label{app:colour_z}

This subsection collects details that are summarised in
Sect.~\ref{sec:redshift} of the main text.

\paragraph{Colour--redshift tracks.}
The $g-r$ colour of galaxies rises from about $0.3$ at $z\approx0$ to
$1.7$--$2.0$ at $z\approx0.4$--$0.6$, and $r-i$ rises from about $0.4$
to about $1.5$ at $z\approx0.7$, as the $4000$\,\AA\ break moves
through the $g$, $r$, and $i$ bands. The sample ends abruptly near
$z\approx1$, which reflects the targeting limits of the SDSS
spectroscopic galaxy samples. In $W1-W2$ the galaxies stay near $0.2$
and show a dip to about $-0.3$ near $z\approx0.6$, which we interpret
as the stellar emission peak near $1.6\,\mu$m in the rest frame moving
through the WISE bands. The colours of quasars remain in a narrow band
($g-r$ and $r-i$ between about $0$ and $0.5$) for
$0.5\lesssim z\lesssim3$, with smooth undulations consistent with
broad emission lines moving through the bands. Above $z\approx3$ the
colour $g-r$ increases (to about $0.6$--$1.0$ at $z\approx3.5$), as
expected when the Lyman-$\alpha$ forest suppresses the $g$-band flux.
The quasar $W1-W2$ is centred above $0.8$ for
$0.3\lesssim z\lesssim3$ and peaks at about $1.5$ near
$z\approx1.5$--$2$. Stars lie on a vertical line at $z\approx0$ that
covers a wide range of colours, because their colours depend on
temperature and not on redshift. Because the colours of galaxies are
monotonic functions of $z$ for $z\lesssim0.7$ and those of quasars are
not, a no-redshift model can estimate the redshift of galaxies from
their colours but not that of quasars.

\begin{figure}[htbp]
\centering
\includegraphics[width=0.45\textwidth]{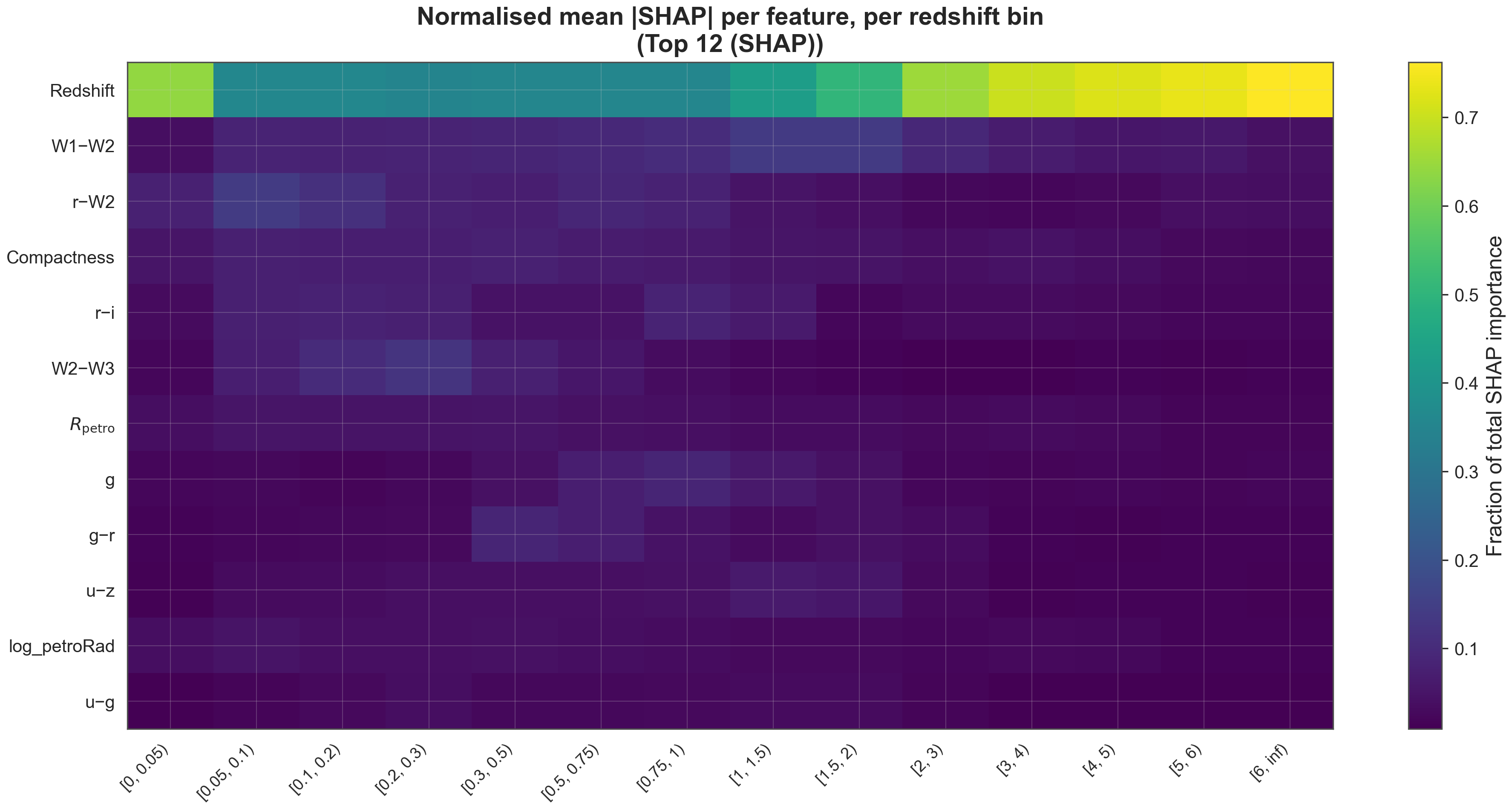}\\[2mm]
\includegraphics[width=0.45\textwidth]{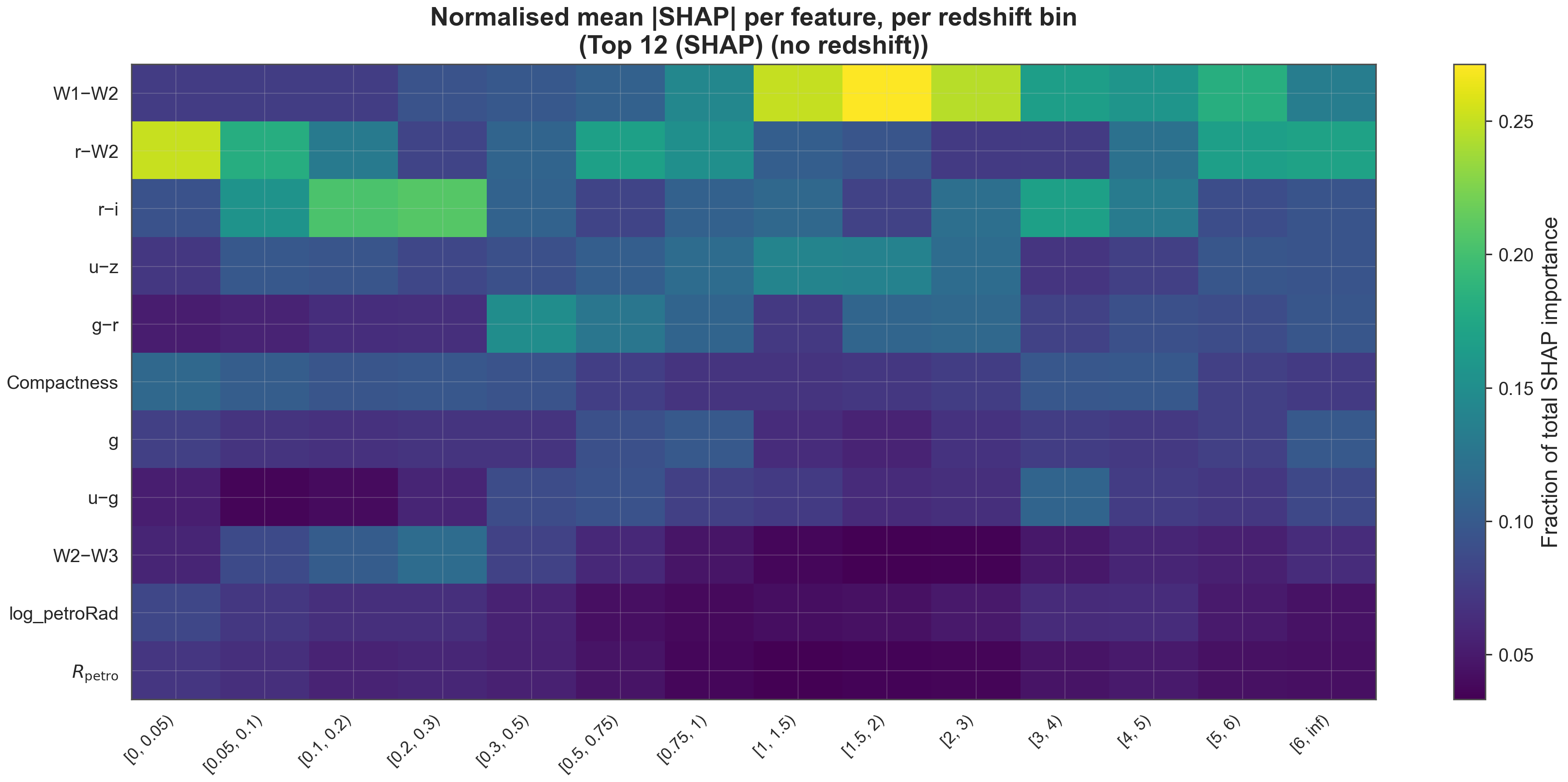}
\caption{Normalised mean absolute SHAP value of each feature in
redshift bins.}
\label{fig:shap_maps}
\end{figure}

\paragraph{SHAP bins.}
The normalised mean $|\mathrm{SHAP}|$ in redshift bins
(Fig.~\ref{fig:shap_maps}) is $0.64$ at $z<0.05$ (where redshift
separates stars from galaxies), has a minimum of $0.36$ at
$0.5\le z<0.75$ (where galaxies and quasars are both abundant and
other features contribute, mainly $g$, $g-r$, and $W1-W2$), and is
$0.76$ at $z\ge4$ (a bin with \num{1366} quasars; see
Table~\ref{tab:recall_binned_all}). Without redshift, the dominant
feature changes with redshift: $r-W2$ at $z<0.05$ (share about
$0.26$), $r-i$ at $0.1\lesssim z\lesssim0.3$, $g-r$ at
$0.3\lesssim z\lesssim0.75$, $W1-W2$ at $1\lesssim z\lesssim3$ (about
$0.25$--$0.27$), and $u-g$ at $3\le z<4$.

\paragraph{Redshift transfer.}
There are no stars at $z\ge0.5$, so the first direction (train low-$z$,
test high-$z$) is a two-class (GALAXY+QSO) problem and its macro-$F_1$
($0.8696$) is not directly comparable with the reverse direction, in
which the model has never seen a star and the STAR $F_1$ is zero by
construction (three-class macro-$F_1 = 0.4208$). For a like-for-like
comparison, the QSO $F_1$ is $0.8187$ when trained at low and tested
at high redshift, and $0.4243$ in the reverse direction. Two effects
contribute: because $z$ is a feature, a tree model evaluated outside
the range seen in training assigns the class of the nearest redshift
leaf, so part of the result is an extrapolation in $z$ itself; and the
QSO population at low redshift, dominated by quasars with a visible
host galaxy, is poorly represented in a high-redshift training set. The
test therefore measures the sensitivity to the redshift range of the
training sample, not a separate physical mechanism.

\subsection{Paired bootstrap and colour diagnostics}
\label{app:paired}

The paired bootstrap of Table~\ref{tab:bootstrap_three_models}
resamples the 11,514 common objects jointly for all four conditions
A--D, so the object-to-object noise that cancels in a paired comparison
does not inflate the interval. The intervals cover the sampling of the
common objects only, not the variance of the training (the models are
held fixed). With only 100 common QSOs, a non-resolved effect is a
statement about power, not about absence of an effect.

The colour diagnostics of Table~\ref{tab:w1_bins} report, for each
colour bin, the Pearson correlation, the Spearman correlation (with a
bootstrap 95\% interval), the $3\sigma$-trimmed Pearson correlation,
the robust scatter ($1.4826\times$ median absolute deviation) of the
DESI-minus-SDSS difference, and the fraction of objects with
$|\Delta{\rm colour}|>1$ mag. The near-zero Pearson correlation of
$W1-W2$ (0.004) is an artefact of heavy tails and of the narrow dynamic
range of the galaxy-dominated sample: the Spearman correlation is 0.72
and the $3\sigma$-trimmed Pearson is 0.88. The agreement degrades with
depth: in quantile bins of SDSS $W1$ the Spearman correlation of
$W1-W2$ peaks at 0.86 in $13.0\le W1<13.5$ and falls to 0.50 in the
faintest common bin ($14.3\le W1<17.8$), and the robust scatter rises
from 0.038 to 0.087 mag over the same range. This is the depth effect
that drives the DESI transfer failure of Sect.~\ref{sec:wise}.

\begin{table}[htbp]
\caption{Stratified bootstrap ranking stability (30 resamples).}
\label{tab:bootstrap}
\centering
\begin{tabular}{lrr}
\hline\hline
Feature & 3-method rank & SHAP rank \\
\hline
Redshift                       & $1.00 \pm 0.00$ & $1.00 \pm 0.00$ \\
$\log_{10}R_{\rm petro}$       & $3.02 \pm 0.37$ & $12.20 \pm 8.73$ \\
Compactness                    & $3.71 \pm 0.36$ & $7.67 \pm 5.86$ \\
$R_{\rm petro}$                & $4.69 \pm 0.43$ & $5.43 \pm 3.72$ \\
$\mu_r$                        & $5.96 \pm 0.59$ & $14.37 \pm 3.21$ \\
$u-z$                          & $6.49 \pm 0.35$ & $12.20 \pm 3.83$ \\
$g-r$                          & $9.84 \pm 0.65$ & $4.80 \pm 1.27$ \\
$W1-W2$                        & $10.18 \pm 0.23$ & $2.17 \pm 0.38$ \\
\hline
\end{tabular}
\end{table}

\begin{table}[htbp]
\caption{Full test-set macro-$F_1$ of the best model across all scenarios, with $95\%$ bootstrap intervals.}
\label{tab:all_scenarios}
\centering
\resizebox{\columnwidth}{!}{%
\begin{tabular}{lrcc}
\hline\hline
Scenario & $n$ feat & Test macro-$F_1$ & 95\% CI \\
\hline
All features (25)               & 25 & 0.9830 & [0.9827, 0.9833] \\
All features (25), no $z$       & 24 & 0.9551 & [0.9546, 0.9556] \\
Top 12 (SHAP)                   & 12 & 0.9834 & [0.9830, 0.9837] \\
Top 6 (3-method)                &  6 & 0.9756 & [0.9752, 0.9760] \\
Top 6 (4-method unified)        &  6 & 0.9784 & [0.9780, 0.9788] \\
Top 6 (SHAP)                    &  6 & 0.9787 & [0.9783, 0.9791] \\
Top 12 (SHAP), no $z$           & 11 & 0.9526 & [0.9520, 0.9532] \\
Top 6 (3-method), no $z$        &  5 & 0.9048 & [0.9039, 0.9057] \\
Top 6 (4-method unified), no $z$ & 5 & 0.9147 & [0.9138, 0.9156] \\
Top 6 (SHAP), no $z$            &  5 & 0.9001 & [0.8992, 0.9010] \\
\hline
\end{tabular}%
}
\end{table}

\begin{table}[htbp]
\caption{Per-class test $F_1$ of the tuned scenarios.}
\label{tab:a1}
\centering
\resizebox{\columnwidth}{!}{%
\begin{tabular}{lrrrrr}
\hline\hline
Scenario & $n$ feat & Macro-$F_1$ & $F_1$(GAL) & $F_1$(QSO) & $F_1$(STAR) \\
\hline
Top 12 (4-method unified) & 12 & 0.9828 & 0.9905 & 0.9594 & 0.9986 \\
Top 12 (SHAP)             & 12 & 0.9834 & 0.9908 & 0.9608 & 0.9987 \\
Top 12 (SHAP), no $z$     & 11 & 0.9526 & 0.9782 & 0.9277 & 0.9518 \\
\hline
\end{tabular}%
}
\end{table}

\begin{table}[htbp]
\caption{Mean test-set $\Delta F_1$ from hyperparameter tuning.}
\label{tab:tuning_gains}
\centering
\begin{tabular}{lrrr}
\hline\hline
Scenario type & RandomForest & XGBoost & LightGBM \\
\hline
Top-12 (with $z$) & $+0.0001$ & $\mathbf{+0.0030}$ & $+0.0012$ \\
Top-6 (with $z$)  & $-0.0009$ & $+0.0006$ & $+0.0007$ \\
Top-12 (no $z$)   & $+0.0003$ & $\mathbf{+0.0034}$ & $\mathbf{+0.0032}$ \\
Top-6 (no $z$)    & $\mathbf{-0.0058}$ & $-0.0002$ & $-0.0007$ \\
\hline
\end{tabular}
\end{table}

\begin{table}[htbp]
\caption{Per-class recall vs redshift bin for three reference scenarios. Dashes indicate bins with no sources of that class.}\label{tab:recall_binned_all}
\centering
\resizebox{\columnwidth}{!}{%
\begin{tabular}{lrrrrrrll}
\hline\hline
$z$ bin & $n_{\rm tot}$ & $n_{\rm GAL}$ & $R_{\rm GAL}$ &
$n_{\rm QSO}$ & $R_{\rm QSO}$ & $n_{\rm STAR}$ & $R_{\rm STAR}$ & scenario \\
\hline
$[0,0.05)$   & 50918  & 18492 & 0.9761 & 93    & 0.2796 & 32333 & 0.9994 & Top-12 SHAP \\
$[0.05,0.1)$ & 41663  & 41399 & 0.9997 & 264   & 0.4886 & 0     & ---    & Top-12 SHAP \\
$[0.1,0.2)$  & 58783  & 57875 & 0.9978 & 908   & 0.7081 & 0     & ---    & Top-12 SHAP \\
$[0.2,0.3)$  & 27785  & 26625 & 0.9946 & 1160  & 0.7672 & 0     & ---    & Top-12 SHAP \\
$[0.3,0.5)$  & 88993  & 85910 & 0.9964 & 3083  & 0.7785 & 0     & ---    & Top-12 SHAP \\
$[0.5,0.75)$ & 129154 & 121455& 0.9952 & 7699  & 0.8352 & 0     & ---    & Top-12 SHAP \\
$[0.75,1)$   & 33312  & 23064 & 0.9777 & 10248 & 0.9265 & 0     & ---    & Top-12 SHAP \\
$[1,1.5)$    & 25884  & 2837  & 0.7032 & 23047 & 0.9800 & 0     & ---    & Top-12 SHAP \\
$[1.5,2)$    & 19021  & 306   & 0.0163 & 18715 & 0.9994 & 0     & ---    & Top-12 SHAP \\
$[2,3)$      & 16699  & 1     & ---    & 16698 & 1.0000 & 0     & ---    & Top-12 SHAP \\
$[3,4)$      & 2893   & 0     & ---    & 2893  & 1.0000 & 0     & ---    & Top-12 SHAP \\
$[4,\infty)$ & 1366   & 0     & ---    & 1366  & 1.0000 & 0     & ---    & Top-12 SHAP \\
\hline
$[0,0.05)$   & 50918  & 18492 & 0.9481 & 93    & 0.2688 & 32333 & 0.9341 & Top-12 SHAP, no $z$ \\
$[0.05,0.1)$ & 41663  & 41399 & 0.9959 & 264   & 0.4886 & 0     & ---    & Top-12 SHAP, no $z$ \\
$[0.1,0.2)$  & 58783  & 57875 & 0.9914 & 908   & 0.7203 & 0     & ---    & Top-12 SHAP, no $z$ \\
$[0.2,0.3)$  & 27785  & 26625 & 0.9804 & 1160  & 0.7914 & 0     & ---    & Top-12 SHAP, no $z$ \\
$[0.3,0.5)$  & 88993  & 85910 & 0.9915 & 3083  & 0.7853 & 0     & ---    & Top-12 SHAP, no $z$ \\
$[0.5,0.75)$ & 129154 & 121455& 0.9898 & 7699  & 0.8367 & 0     & ---    & Top-12 SHAP, no $z$ \\
$[0.75,1)$   & 33312  & 23064 & 0.9516 & 10248 & 0.9296 & 0     & ---    & Top-12 SHAP, no $z$ \\
$[1,1.5)$    & 25884  & 2837  & 0.8195 & 23047 & 0.9209 & 0     & ---    & Top-12 SHAP, no $z$ \\
$[1.5,2)$    & 19021  & 306   & 0.6797 & 18715 & 0.9650 & 0     & ---    & Top-12 SHAP, no $z$ \\
$[2,3)$      & 16699  & 1     & ---    & 16698 & 0.9780 & 0     & ---    & Top-12 SHAP, no $z$ \\
$[3,4)$      & 2893   & 0     & ---    & 2893  & 0.9115 & 0     & ---    & Top-12 SHAP, no $z$ \\
$[4,\infty)$ & 1366   & 0     & ---    & 1366  & 0.3865 & 0     & ---    & Top-12 SHAP, no $z$ \\
\hline
\end{tabular}%
}
\end{table}

\begin{table}[htbp]
\caption{Per-feature permutation importance.}
\label{tab:perm_standard_v2}
\centering
\resizebox{\columnwidth}{!}{%
\begin{tabular}{lrr}
\hline\hline
Feature & Importance & Std \\
\hline
Redshift              & 0.5307 & 0.0021 \\
$g-r$                 & 0.0205 & 0.0007 \\
$r-i$                 & 0.0201 & 0.0004 \\
$\mu_r$               & 0.0187 & 0.0008 \\
$r-W2$                & 0.0093 & 0.0002 \\
$W1-W2$               & 0.0067 & 0.0001 \\
$u-g$                 & 0.0055 & 0.0005 \\
$i-z$                 & 0.0009 & 0.0002 \\
Compactness           & 0.0005 & 0.0002 \\
$u-z$                 & 0.0003 & 0.0001 \\
$\log_{10}R_{\rm petro}$ & 0.0001 & 0.0001 \\
$R_{\rm petro}$       & 0.0001 & 0.0001 \\
\hline
\end{tabular}
}
\end{table}

\begin{table}[htbp]
\caption{Spearman correlation between the redshift and four colours,
by class (training set).}
\label{tab:spearman}
\centering
\resizebox{\columnwidth}{!}{%
\begin{tabular}{lccc}
\hline\hline
Colour & GALAXY & QSO & STAR \\
\hline
$r-W2$ & $+0.886$ & $-0.420$ & $+0.140$ \\
$i-W1$ & $+0.877$ & $-0.500$ & $+0.140$ \\
$r-i$  & $+0.855$ & $-0.042$ & $+0.154$ \\
$z-W1$ & $+0.849$ & $-0.500$ & $+0.137$ \\
\hline
\end{tabular}
}
\end{table}

\begin{table}[htbp]
\caption{Cross-redshift transfer test (with $z$ included as a feature).}
\label{tab:redshift_transfer}
\centering
\resizebox{\columnwidth}{!}{%
\begin{tabular}{lrrrrr}
\hline\hline
Scenario & $n_{\rm train}$ & $n_{\rm test}$ & macro-$F_1$ & QSO $F_1$ & STAR $F_1$ \\
\hline
Train low-$z$, test high-$z$ & 1,491,233 & 1,065,384 & 0.8696$^*$ & 0.8187 & 0.0000 \\
Train high-$z$, test low-$z$ & 1,065,384 & 1,491,233 & 0.4208 & 0.4243 & 0.0000 \\
Random split (reference)     & 2,556,617 & 547,847 & 0.9834 & 0.9608 & 0.9987 \\
\hline
\multicolumn{6}{l}{\footnotesize $^*$Two-class (GAL+QSO) macro-$F_1$, because no stars are present at $z\ge0.5$.}
\end{tabular}
}
\end{table}

\begin{table}[htbp]
\caption{Magnitude-shift test ($r<20$ versus $r\ge20$); macro-$F_1$
averaged over the two transfer directions.}
\label{tab:cross_z}
\centering
\resizebox{\columnwidth}{!}{%
\begin{tabular}{lrrr}
\hline\hline
Scenario & In regime & Across regimes & Drop \\
\hline
Top 12 (SHAP), with $z$ & 0.984 & 0.975 & $-0.010$ \\
Top 12 (SHAP), no $z$   & 0.933 & 0.851 & $-0.082$ \\
\hline
\end{tabular}
}
\end{table}

\begin{table}[htbp]
\caption{DESI DR1 sample composition and redshift distribution.}
\label{tab:desi_sample}
\centering
\resizebox{\columnwidth}{!}{%
\begin{tabular}{lrrrrr}
\hline\hline
Class & Count & Fraction & $z_{\rm median}$ & $z_{16}$ & $z_{84}$ \\
\hline
GALAXY & 226,967 & 0.986 & 0.319 & 0.118 & 0.835 \\
QSO    &   2,421 & 0.0105 & 0.564 & 0.240 & 1.432 \\
STAR   &     847 & 0.0037 & 0.000 & 0.000 & 0.000 \\
\hline
\end{tabular}
}
\end{table}

\begin{table}[htbp]
\caption{QSO purity at a completeness of $0.90$ on DESI DR1 for
assumed quasar fractions of $1\%$ and $5\%$. The values correspond to a
false-positive rate of $0.573$ at that completeness.}
\label{tab:desi_purity}
\centering
\resizebox{\columnwidth}{!}{%
\begin{tabular}{llrr}
\hline\hline
Prior & Class & Completeness & Purity \\
\hline
DESI (1\% QSO) & QSO & 0.90 & 0.0156 \\
DESI (5\% QSO) & QSO & 0.90 & 0.0763 \\
\hline
\end{tabular}
}
\end{table}

\begin{table}[htbp]
\caption{Per-class recall vs redshift bin on the \num{230235} DESI DR1
sources for the 10-feature SDSS-trained model (with $z$).}
\label{tab:desi_recall_binned}
\centering
\resizebox{\columnwidth}{!}{%
\begin{tabular}{lrrrrrr}
\hline\hline
$z$ bin & $n_{\rm tot}$ & $n_{\rm GAL}$ & $R_{\rm GAL}$ &
$n_{\rm QSO}$ & $R_{\rm QSO}$ & $n_{\rm STAR}$ ($R_{\rm STAR}$) \\
\hline
$[0, 0.5)$   & 154448 & 152710 & 0.994 & 1112 & 0.308 & 626 (1.000) \\
$[0.5, 1)$   &  55606 &  54806 & 0.944 &  800 & 0.375 & --- \\
$[1, 1.5)$   &  17438 &  17282 & 0.288 &  128 & 0.867 & --- \\
$[1.5, 2)$   &   2070 &   2070 & 0.000 &    0 & ---   & --- \\
$[2, \infty)$&     673 &      0 & ---   &  673 & 1.000 & --- \\
\hline
\end{tabular}
}
\end{table}

\begin{table}[htbp]
\caption{Simplified reproductions of three published methods on the
fixed test set of this paper, with the same default Random Forest and
the same train/validation/test split. The feature sets are our best
reconstruction of each method's feature list, not the original
pipelines.}
\label{tab:lit_baselines}
\centering
\resizebox{\columnwidth}{!}{%
\begin{tabular}{llrr}
\hline\hline
Method & Feature set & Macro-$F_1$ & QSO $F_1$ \\
\hline
\multicolumn{4}{l}{\emph{With spectroscopic redshift}} \\
Bai et al.\ 2019     & 9 colours (optical+WISE+2MASS) & 0.9810 & 0.9555 \\
Clarke et al.\ 2020  & optical+WISE+morphology        & 0.9830 & 0.9585 \\
Lourens et al.\ 2024 & DR+kNN features                & 0.9834 & 0.9614 \\
\hline
\multicolumn{4}{l}{\emph{Without spectroscopic redshift}} \\
Bai et al.\ 2019     & 9 colours (optical+WISE+2MASS) & 0.9525 & 0.9229 \\
Clarke et al.\ 2020  & optical+WISE+morphology        & 0.9545 & 0.9254 \\
Lourens et al.\ 2024 & DR+kNN features                & 0.9551 & 0.9277 \\
\hline
\end{tabular}
}
\end{table}


\end{appendix}

\end{document}